**Dr. Petar Radanliev**

Parks Road,

Oxford OX1 3PJ

United Kingdom

Email: petar.radanliev@cs.ox.ac.uk

Phone: +389(0)79301022

BA Hons., MSc., Ph.D. Post-Doctorate

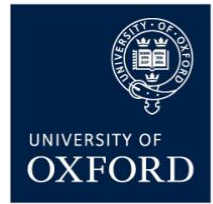

DEPARTMENT OF
COMPUTER
SCIENCE

UNIVERSITY OF OXFORD

# SBOMs into Agentic AIBOMs: Schema Extensions, Agentic Orchestration, and Reproducibility Evaluation

*Corresponding author: Petar Radanliev[1]: Email: petar.radanliev@cs.ox.ac.uk*

Petar Radanliev*

Department of Computer Sciences, University of Oxford, Wolfson Building, Parks Rd, Oxford OX1 3QG; The Alan Turing Institute, British Library, 96 Euston Rd., London NW1 2DB; Email: * petar.radanliev@cs.ox.ac.uk

**Other IDs**

ORCID: https://orcid.org/0000-0001-5629-6857

ResearcherID: L-7509-2015

ResearcherID: M-2176-2017

Scopus Author ID: 57003734400

Loop profile: 839254

ResearcherID: L-7509-2015

Carsten Maple, The Alan Turing Institute, British Library, 96 Euston Rd., London NW1 2DB, University of Warwick – WMG, 6 Lord Bhattacharyya Way, Coventry CV4 7AL

Omar Santos, Cisco Systems, RTP, North Carolina, United States

Kayvan Atefi, Computer Science, School of Digital & Physical Sciences, Faculty of Science and Engineering, University of Hull, Hull, United Kingdom

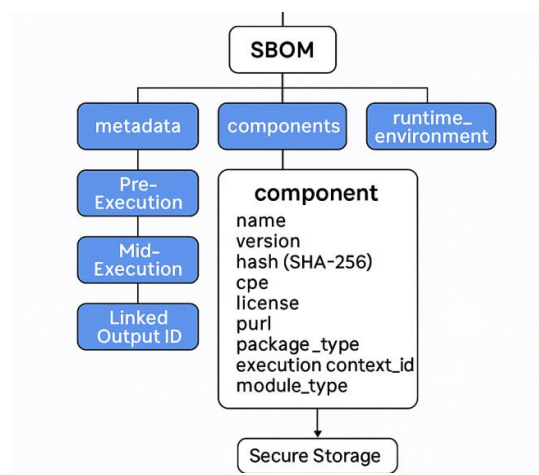

**Dr. Petar Radanliev**

Parks Road,

Oxford OX1 3PJ

United Kingdom

Email: petar.radanliev@cs.ox.ac.uk

Phone: +389(0)79301022

BA Hons., MSc., Ph.D. Post-Doctorate

DEPARTMENT OF
# COMPUTER SCIENCE

UNIVERSITY OF OXFORD

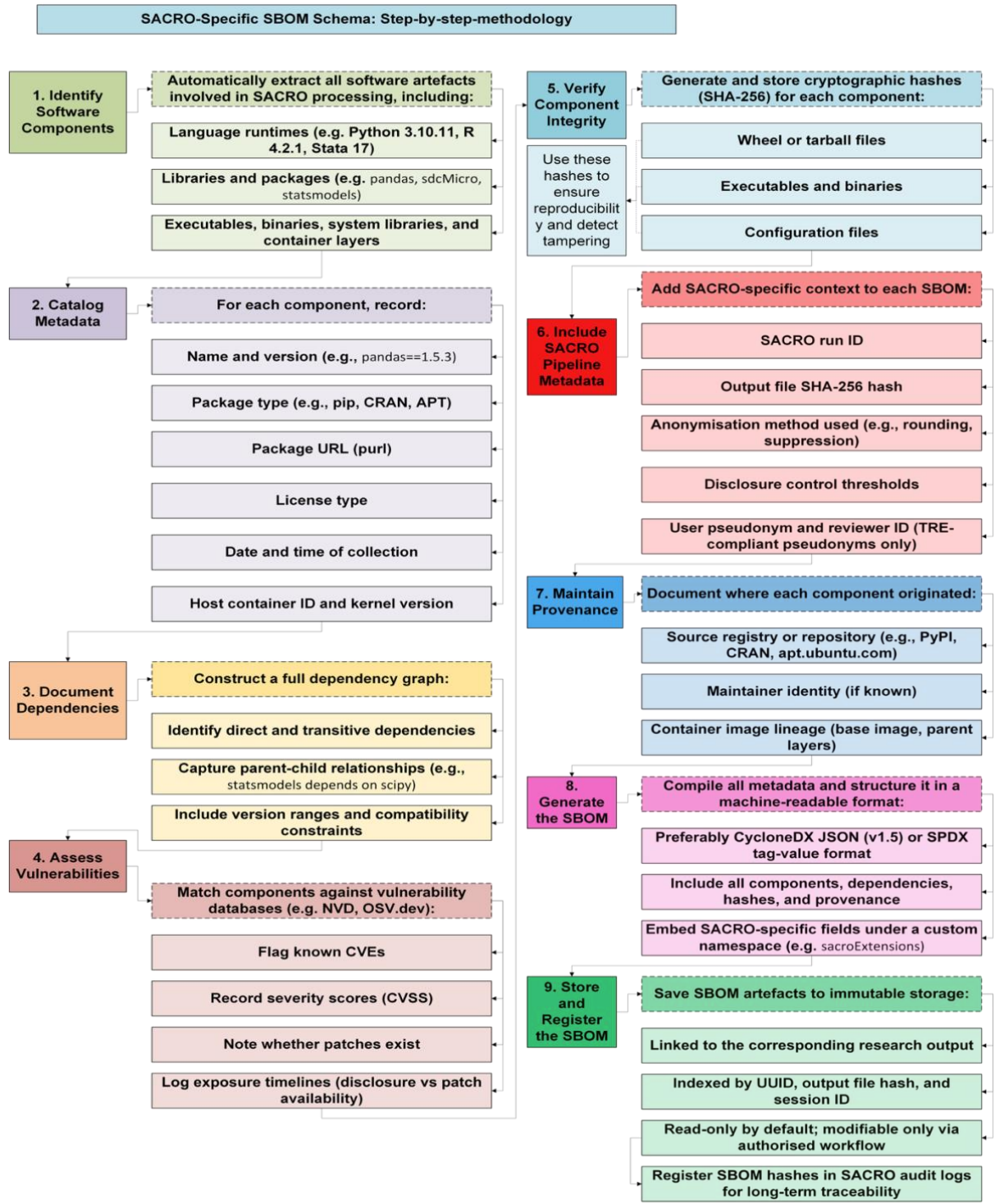

**Abstract**: Software supply-chain security requires provenance mechanisms that support reproducibility and vulnerability assessment under dynamic execution conditions. Conventional Software Bills of Materials (SBOMs) provide static dependency inventories but cannot capture runtime behaviour, environment drift, or exploitability context. This paper

**Dr. Petar Radanliev**
Parks Road,
Oxford OX1 3PJ
United Kingdom
Email: petar.radanliev@cs.ox.ac.uk
Phone: +389(0)79301022
BA Hons., MSc., Ph.D. Post-Doctorate

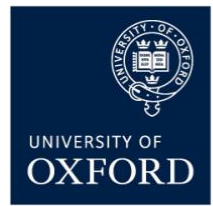

introduces agentic Artificial Intelligence Bills of Materials (AIBOMs), extending SBOMs into active provenance artefacts through autonomous, policy-constrained reasoning. We present an agentic AIBOM framework based on a multi-agent architecture comprising (i) a baseline environment reconstruction agent (MCP), (ii) a runtime dependency and drift-monitoring agent (A2A), and (iii) a policy-aware vulnerability and VEX reasoning agent (AGNTCY). These agents generate contextual exploitability assertions by combining runtime execution evidence, dependency usage, and environmental mitigations with ISO/IEC 20153:2025 Common Security Advisory Framework (CSAF) v2.0 semantics. Exploitability is expressed via structured VEX assertions rather than enforcement actions. The framework introduces minimal, standards-aligned schema extensions to CycloneDX and SPDX, capturing execution context, dependency evolution, and agent decision provenance while preserving interoperability. Evaluation across heterogeneous analytical workloads demonstrates improved runtime dependency capture, reproducibility fidelity, and stability of vulnerability interpretation compared with established provenance systems, with low computational overhead. Ablation studies confirm that each agent contributes distinct capabilities unavailable through deterministic automation.

## 1 INTRODUCTION

Software supply-chain security has moved beyond static transparency artefacts towards active, context-aware provenance systems capable of supporting continuous vulnerability assessment, reproducibility, and policy-aligned assurance. Conventional Software Bills of Materials (SBOMs) provide a machine-readable inventory of components, but they remain fundamentally descriptive: they enumerate dependencies without reasoning about how those dependencies are instantiated, executed, mitigated, or rendered exploitable within a concrete runtime environment. As software systems increasingly rely on dynamic loading, late binding, federated services, and autonomous orchestration, static SBOMs are no longer sufficient to support trustworthy exploitability analysis or reproducible security assurance.

In parallel, recent advances in agentic AI have demonstrated that autonomous, policy-constrained agents can perform structured reasoning under uncertainty, coordinate across distributed observability layers, and adapt decisions in response to runtime evidence. These capabilities directly address long-standing limitations in SBOM-centric security tooling: the inability to detect environment drift, to distinguish presence from exploitability, and to bind vulnerability interpretation to execution context rather than package metadata alone. This convergence motivates a shift from SBOMs as passive inventories to agentic Artificial Intelligence Bills of Materials (AIBOMs): provenance artefacts that integrate software composition data with autonomous reasoning, runtime telemetry, and standards-aligned vulnerability semantics.

This paper formalises that shift. We introduce an agentic AIBOM framework that extends existing SBOM standards with (i) execution-context awareness, (ii) dynamic dependency evolution, and (iii) auditable decision traces produced by interacting agents. The framework employs a multi-agent architecture in which specialised agents operate over distinct perception spaces and decision policies: a baseline environment reconstruction agent, a runtime drift and dependency-context agent, and a policy-aware exploitability reasoning agent. In this work, "agentic AI" refers to autonomous software components characterised by (i) a bounded perception space, (ii) internal state, (iii) explicit decision policies, and (iv) structured interaction with other agents or system components. The agents described here are not learning-based or generative systems, but policy-constrained decision modules designed for auditable reasoning under uncertainty. Rather than replacing SBOM standards, these agents augment SBOMs with operational intelligence, transforming them into active cybersecurity objects capable of issuing contextual vulnerability and reproducibility assertions.

A critical enabler of this transformation is the recent ratification of ISO/IEC 20153:2025, which elevates the OASIS Common Security Advisory Framework (CSAF) v2.0 to an international standard. CSAF provides a schema-stable, regulator-recognised

**Dr. Petar Radanliev**

Parks Road,

Oxford OX1 3PJ

United Kingdom

Email: petar.radanliev@cs.ox.ac.uk

Phone: +389(0)79301022

BA Hons., MSc., Ph.D. Post-Doctorate

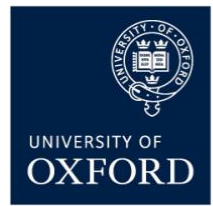

structure for vulnerability advisories, remediation guidance, and VEX-aligned exploitability statements. By anchoring agentic reasoning outputs to CSAF semantics, AIBOMs can express exploitability determinations in a form that is machine-verifiable, interoperable, and auditable, avoiding ad-hoc or opaque risk scoring. Importantly, this work does not assume that all vulnerabilities can or should be automatically gated; instead, it treats VEX assertions as evidence-bearing statements whose validity depends on runtime conditions and documented mitigations.

The proposed AIBOM schema extends established SBOM formats (CycloneDX, SPDX) with fields capturing runtime environment state, dependency resolution timelines, agentic decision provenance, and CSAF-linked advisory evidence. These extensions are deliberately minimal and compositional: they preserve compatibility with existing SBOM generators and consumers while enabling higher-order reasoning that static inventories cannot support. The resulting artefact binds software composition, execution context, and vulnerability interpretation into a single, cryptographically verifiable provenance object. While the framework is applicable to a broad class of regulated and high-assurance software pipelines, we evaluate it in controlled analytic environments where reproducibility and auditability are mandatory. In such settings, unobserved software drift or misinterpreted vulnerabilities can invalidate outputs even when data and code remain unchanged. These environments therefore provide a stringent testbed for assessing whether agentic AIBOMs deliver measurable improvements over deterministic automation and post-hoc provenance capture.

The contributions of this paper are threefold. First, we define the conceptual and technical requirements for transforming SBOMs into agentic AIBOMs, grounded in distributed AI principles and contemporary vulnerability standards. Second, we present a concrete multi-agent architecture and schema extensions that operationalise these requirements while remaining standards-aligned. Third, through benchmarking and ablation studies, we demonstrate that agentic AIBOMs improve runtime dependency fidelity, reproducibility accuracy, and contextual exploitability determination compared with established provenance systems, without imposing prohibitive computational overhead. To address the limitation of testing and practical application, this paper proposes the integration of a structured SBOM schema [1] into the open-source data on the Agentic AIBOM orchestration pipeline. The SBOM, as defined here, is a machine-readable inventory [2], that captures the complete set of software artefacts present in the analytical environment at the time of output generation [3]. This includes exact versions of statistical packages, disclosure control libraries, language runtimes (e.g., R, Python, Stata), container metadata, system-level dependencies, and any executable scripts loaded during runtime [4].

By integrating SBOM provenance, agentic AI, and ISO-standardised advisory semantics, this work establishes a reproducible and auditable foundation for next-generation software supply-chain assurance. It reframes the bill of materials into an active, reasoning-capable artefact suited to the realities of modern, dynamic software systems.

## 2   LITERATURE REVIEW - WHAT ARE THE SBOM STANDARDISED ONTOLOGIES?

SBOM is a detailed inventory that records all software product components. It is crucial for identifying vulnerable components across the entire software supply chain and aiding asset owners in decision-making when a vulnerable product is identified. SBOM is designed for automation, helping identify, track, and manage software components, particularly regarding their security aspects. As an '*example of the data representation',* SBOMs may include libraries, packages, modules, and their respective versions, licenses, and source locations. This can be represented in a machine-readable format, such as JSON or XML, outlining each component's details.

Software Supply Chain Cyber Risk refers to the diverse components and processes used in software development and distribution. The Cyber Supply Chain Management and Transparency Act of 2014 transposed the Bill of Materials (BOM) concept from traditional manufacturing to software. Subsequent legislative acts and executive orders mandated SBOMs for governmental software procurement. The role of SBOM is to list the "ingredients" that constitute software, offering a machine-readable inventory that helps manage supply chain assets and vulnerabilities. While the complexity of the ecosystem precludes a universal SBOM solution, automation is critical to harnessing its full potential for cybersecurity transparency.

We organise prior work along three threads: (i) SBOM standards and generators (CycloneDX, SPDX; generation accuracy and consumption quality), (ii) Provenance and reproducibility frameworks (packaging, unit-level validation, PROV graphs),

**Dr. Petar Radanliev**
Parks Road,
Oxford OX1 3PJ
United Kingdom
Email: petar.radanliev@cs.ox.ac.uk
Phone: +389(0)79301022
BA Hons., MSc., Ph.D. Post-Doctorate

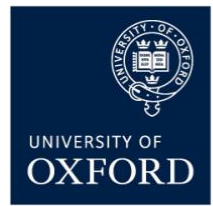

and (iii) Policy and governance mappings (CSAF/VEX, GDPR/NIST/ISO alignment). We use this structure to expose the specific gap our method addresses: turning SBOMs into machine-verifiable compliance and reproducibility artefacts for regulated analytic workflows, rather than static inventories or developer-centric artefacts.

Although SBOMs can identify potential vulnerabilities, not all are exploitable, and when it gets to cybersecurity fixes [5], the debate on who is responsible for patches and updates is a big problem [6]. Some solutions discussed in existing literature include replacing the Common Vulnerability Scoring System Calculator (CVSS), the Base Score, the Temporal Score and the Environmental Score [7] with three possible strategies based on the vulnerability score only: a). patch now; b). patch later; and c). accept the risk [8]. Given the volume of requests and updates, manual patching and updates daily is impossible, especially because there is also a timeframe of concern, which requires automation process, however, some advisories are still written in a text editor and a human-readable format [9]. The contributions of this paper enable enhanced tracking of dependencies, the new process ensuring reproducibility, and supports audits and compliance procedures.

## 2.1    Summary of ontologies for standardisation

There are standardised tools and ontologies to enhance security information exchange and automate vulnerability management, such as the 'Reference Ontology for Cybersecurity Operational Information' [10]. Another Ontology is the CYBEX framework [11], which represents a significant step in establishing a global standard for exchanging cybersecurity information. As the International Telecommunication Union Telecommunication Standardization Sector (ITU-T) initiative, CYBEX aims to standardise how cybersecurity entities communicate and ensure the integrity of this exchange.

CYBEX's architecture facilitates the accumulation, discovery, querying, assurance, and transportation of cybersecurity information. It controls 18 existing specifications alongside three newly created ones, effectively building on the foundation of de facto standards for enhanced compatibility and utility. The Information Description block uses established specifications like CVE and CWE to systematically describe and structure cybersecurity information. By enabling the seamless integration of these specifications, CYBEX adopts an ecosystem where knowledge from various sources can be shared and accessed globally, enabling resource-constrained entities to participate equally in cybersecurity efforts.

To identify peer reviewed ontologies for standardisation, we searched the Web of Science Core Collection and identified one study on building resilient medical technology supply chains with a software bill of materials [12], and a second study on the topic of codifying attack surfaces [13]. Given the limited results, we continued to investigate the related topic of software supply chain risk. We expanded into the topic of software supply chain risk, which can be defined as the collection of components, libraries, tools, and processes used to develop, build, and publish the software. In supply chain manufacturing, there is a well-established concept of a 'bill of materials' (BOM) [14]. More recent developments apply the same principles to software supply chains. The 'Cyber Supply Chain Management and Transparency Act of 2014' [15] proposed that US government agencies obtain SBOMs for all new software. This led to the 'Internet of Things Cybersecurity Improvement Act of 2017' [16] and, 'The US Executive Order on Improving the Nation's Cybersecurity of May 12, 2021 [17] ordered The National Institute of Standards and Technology (NIST) to issue guidance on '*providing a purchaser a Software Bill of Materials (SBOM) for each product.'*

SBOM can be defined as a nested (machine-readable) inventory of software, a list of ingredients that make up software components and dependencies, and their hierarchical relationships [18], [19]. However, because of '*the diverse needs of the software ecosystem, there is no one-size-fits-all solution'* [20], and the problem with sharing and exchanging is that *'To fully realize the benefits of SBOMs and software component transparency, machine processing and automation are necessary'* [20]. SBOM enables us to identify potentially vulnerable components, but a vulnerability associated with a software component is not necessarily exploitable. At the time of writing this paper (June 2025), the CVE index contained over 191,633 vulnerabilities [21], with 32,760 new vulnerabilities published in 2022 and around 22,000 published in 2021. The CVE ecosystem now (February 2026), after peer review, contains well over two hundred thousand disclosed vulnerabilities, with annual publication rates exceeding thirty thousand entries, rendering exhaustive vulnerability response infeasible without contextual filtering. Organisations cannot perform cybersecurity risk management for all known vulnerabilities, and risk management is based on

**Dr. Petar Radanliev**
Parks Road,
Oxford OX1 3PJ
United Kingdom
Email: petar.radanliev@cs.ox.ac.uk
Phone: +389(0)79301022
BA Hons., MSc., Ph.D. Post-Doctorate

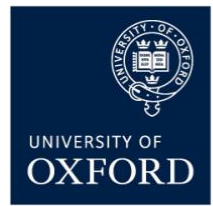

their cyber risk tolerance, which is based on the likelihood and severity (frequency and magnitude) of the risk materialising. If a vulnerability is not assessed for exploitability, it is impossible to predict likelihood. Just one Windows vulnerability opened the door for WannaCry ransomware attacks via the EternalBlue exploit, while the Mirai botnet spread by exploiting multiple vulnerabilities. From this, we can conclude that all vulnerabilities need to be risk assessed, and according to CVE, almost 11% of all vulnerabilities can be categorised as 'critical' [22] because they enable hackers to compromise the apps and data of users of the same hardware. In early 2020, the first critical vulnerabilities in a major cloud infrastructure were discovered - disproving '*the assumption that cloud infrastructures are secure*' [23]. Vulnerabilities also need to be risk assessed because around 75% of attacks in 2020 used vulnerabilities that were at least two years old and high-risk vulnerabilities are present on the network perimeters of 84% of companies [24].

Determining the value of SBOMs, for increasing the visibility of vulnerabilities, is not difficult, but it adds value to producers and vendors, not end users, because end users pay for a product that is sold as inherently secure. Hence, the real difficulty is to determine the value of SBOMs for end users. One specific use case is the automated use of SBOMs for CVE (vulnerability) management. Although there are a few tools designed to manage the software development process of vulnerability management, what is missing is '*easy-to-use and low cost tools and third party services… passing vulnerability data to the vulnerability and configuration management tools that are now deployed by end users – is currently being addressed by nobody*' [25]. The most useful tool that could be found in open source at present, that ingests, analyses, monitors, and produces real-time intelligence reports, is the Dependency-Track [26].

## 2.2 Automation and scaling requirements for processing SBOMs

One of the leading reasons why software users are not requesting SBOMs is that approximately 95% of all vulnerabilities for components listed in an SBOM won't be exploitable in the product [27]. If users start calling suppliers to check each component vulnerability manually, around 95% of that time will be wasted. Since cybersecurity teams are already overstretched by responding to exploitable vulnerabilities detected by their scanners, checking 95% of non-exploitable vulnerabilities doesn't seem like a good idea. A glimpse of the required volume of checks can be seen from correspondence between the creator of the Dependency-Track tool and the Sonatype staff regarding the tool's request for their OSS Index [28] of open-source component vulnerabilities and dependencies. In a response on 25th of September 2022 [29], Sonatype reported '202 million requests to the OSS Index, each request for up to 100 components, and the average software product listed in the OSS Index includes 135 components. This is close to the average software product in SBOM, reportedly about 150 third-party components. But for simplicity, let's keep the figure to 202 million requests for 100 components. This produces about 20 billion instances requested for vulnerability components in 30 days, and this is only for the Dependency-Track tool requests to the OSS Index. This presents a complex challenge for securing the software supply chain, especially because modern software applications heavily rely on diverse third-party components, libraries, and frameworks sourced from various vendors and open-source repositories [30].

Building upon the earlier stated number of requests from the Dependency-Track tool to the OSS Index, the next OSS Index reported an increase from 202 million to 270 million requests per month. If the requests continue to increase at that rate, the number will double every year and a half. To determine how many organisations used the Dependency-Track tool, the creator calculated that each organisation, on average, makes around 250 requests per day, and the number of 202 million is divided over 30 days; the estimated number of organisations using the tool is 27,000. Since the Dependency-Track tool is not the only tool, and the OSS Index is not the only database, we can expect at least 50,000 organisations to use SBOMs for vulnerability management. Most of these organisations are software developers seeking to patch vulnerabilities in the components of their products or replace vulnerable components not patched on time by the component supplier. Just one simple example used by hackers is the attempt to backdoor SSH via an infiltration operation in one of its dependencies, namely the XZ library, which explains the importance of robust software supply chain security [31]. This requires an accurate and compliant SBOM Generator with incremental construction, because '*existing SBOM generators can not compose a complete SBOM with information that developers know best and entries hidden in the dependencies' metadata in one go*' [32].

**Dr. Petar Radanliev**
Parks Road,
Oxford OX1 3PJ
United Kingdom
Email: petar.radanliev@cs.ox.ac.uk
Phone: +389(0)79301022
BA Hons., MSc., Ph.D. Post-Doctorate

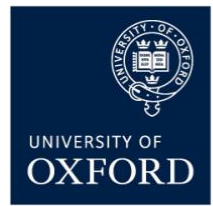

## 2.3 From Static SBOMs to Agent-Mediated Provenance Systems

Recent work on provenance, runtime observability, and autonomous system governance suggests a shift from static metadata capture towards decision-centric artefacts that encode not only system state but the rationale for system actions. However, this literature remains largely disconnected from SBOM standardisation and vulnerability governance. The present work bridges this gap by embedding agentic decision logic directly into the bill-of-materials lifecycle, transforming SBOMs into AIBOMs capable of expressing context, evidence, and bounded reasoning.

## 3 METHODOLOGIES

The methodology is grounded on established benchmark selection and fairness. We benchmark against ReproZip, SciUnit, and ProvStore because they represent the principal design families in computational provenance: packaging for replay, unit-level scientific validation, and PROV metadata graphs, respectively. Where tools optimise for different goals, we report complementary strengths and avoid like-for-like claims outside their design intent. In this paper, policy refers to machine-interpretable constraints governing execution, disclosure, and vulnerability handling within a deployment context. These policies may be institutional (e.g., disclosure thresholds), technical (e.g., sandbox restrictions), or procedural (e.g., reviewer escalation rules), but are treated uniformly as external inputs to the agentic AIBOM reasoning process.

To ensure reproducibility semantics, we distinguish exact parity (byte-identical artefacts verified by SHA-256) from semantic parity (task-specific equivalence within a pre-specified tolerance $\varepsilon$). For statistical outputs, semantic parity requires identical row counts and disclosure suppression ratios with numerical differences bounded by $\varepsilon=1e-12$ for deterministic routines and $\varepsilon=1e-6$ for floating-point ML workloads. Warnings that do not affect output artefacts are logged but not counted as failures. We report both rates and treat semantic-only matches as *benign drift* while any breach of policy thresholds (e.g., SDC failure) is a critical divergence.

To ensure dynamic dependency accuracy we evaluate capture accuracy for declared vs runtime-implicit dependencies. Ground truth is constructed by instrumenting package managers and import hooks to record all loaded modules and shared libraries. For each workload we compute Capture Rate, FPR, and FNR by comparing generated SBOMs (Syft, CycloneDX 1.5; SPDX 2.3) against ground truth. We additionally measure latency to completeness across the pre/mid/post snapshots. The architectural contributions, agent-mediated provenance context, contextual VEX assertion generation, and standards-aligned AIBOM schemas, are independent of any specific governance platform.

## 3.1 Fault-Tolerance and Reliability Considerations in Agent-Assisted SBOM Capture

Although automated agents significantly improve the coverage and temporal accuracy of SBOM capture within an agentic AIBOM pipeline, their introduction increases architectural complexity and introduces additional failure modes. To ensure operational reliability under adversarial or resource-constrained conditions, the orchestration layer incorporates explicit fault-tolerance mechanisms designed to prevent partial, inconsistent, or stale SBOM artefacts from being used in disclosure decisions. First, each agent, MCP for pre-execution capture, A2A for runtime telemetry, and AGNTCY for policy binding, executes under a strict health-check regime that records heartbeat signals, resource utilisation, capture completeness, and deviation from expected dependency profiles. Any anomalous state (e.g., silent termination, extended inactivity, malformed SBOM fragments, or missing dependency chains) triggers fail-closed behaviour: the execution environment halts output release, records a structured incident event, and surfaces a reproducible SBOM-diff diagnostic to the reviewer interface. Secondly, each capture stage is independently verifiable through cross-snapshot reconciliation. The orchestration engine performs deterministic merging of pre-, mid-, and post-execution snapshots, rejecting inconsistent merges and requiring an agent resynchronisation cycle before proceeding.

To mitigate the risk of agent-induced corruption or tampering, all SBOM fragments are signed using session-scoped keys stored in a protected execution enclave, and the composite hash over {SBOM, script, config, outputs} is recomputed after every stage transition. This prevents an agent failure from silently altering the provenance chain or suppressing vulnerability information. Runtime dependency enumeration is further validated by cross-referencing import-hook telemetry with OS-level

**Dr. Petar Radanliev**
Parks Road,
Oxford OX1 3PJ
United Kingdom
Email: petar.radanliev@cs.ox.ac.uk
Phone: +389(0)79301022
BA Hons., MSc., Ph.D. Post-Doctorate

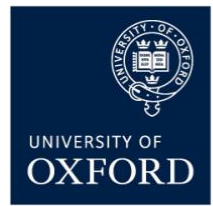

package metadata to detect omissions caused by agent misclassification or transient container-layer faults. If reconciliation thresholds are breached, for example, if >2% of expected components are missing or incorrectly hashed, the system issues an integrity-violation event and requires human adjudication. These design measures ensure that agentic orchestration enhances automation without undermining the determinism and auditability required for high-assurance reproducibility and security governance.

### 3.2      Formalising Agentic Components and Contextual VEX Integration

The proposed agentic AIBOM framework employs three autonomous components—MCP, A2A, and AGNTCY—that operate as semi-independent decision modules responsible for environment reconstruction, runtime dependency reasoning, and policy-aligned exploitability assertions. Their design follows principles from distributed artificial intelligence and multi-agent systems, where an "agent" is characterised by (i) a well-defined perception space, (ii) internal state transitions, (iii) local decision policies, and (iv) message-based interaction with other components.

**Agentic Structure and Decision Model.**

*Perception.*

Each agent consumes a specific subset of runtime observables: MCP ingests container metadata and initial dependency states; A2A monitors dynamic imports, late-bound modules, and CVE feed deltas; AGNTCY consumes policy rules, VEX updates, and policy-defined thresholds.

*Local State.*

The agents maintain internal working memories: MCP tracks unresolved dependencies and unresolved SBOM fragments; A2A maintains a rolling window of runtime telemetry and environment-drift metrics; AGNTCY stores policy contexts, mitigation metadata, and exploitability assertions.

*Decision Policies.*

Each agent maps perceived events to a constrained action set. MCP autonomously determines whether a pre-execution SBOM is complete or requires additional probing. A2A issues environment-drift alerts when live telemetry diverges from expected dependency graphs. AGNTCY determines whether a vulnerability observed by A2A is "Relevant", "Irrelevant", or "Requires escalation" given the runtime environment, available mitigations, and component usage. These decisions produce structured messages interpreted by succeeding phases of the AIBOM-enabled workflow.

*Coordination.*

Agent interaction is mediated through schema-constrained messages rather than shared memory. This design enforces role separation: the correctness of MCP's baseline capture does not depend on A2A's behaviour, and AGNTCY's VEX decision logic can be audited independently of SBOM generation.

Taken together, these characteristics satisfy the operational definition of agentic orchestration in secure analytic workflows: autonomy over local decisions, explicit policies, internal state transitions, and structured inter-agent communication.

**Granular VEX Semantics and Current Implementation Boundaries.**

VEX integration is frequently misunderstood as producing final exploitability determinations, whereas its formal role is to provide *contextual assertions* about whether a vulnerability is exploitable in a specific operational environment. To distinguish scope from future work, the evaluated AIBOM implementation supports three levels of VEX granularity:

1.   **Presence Assertions (implemented).**

Each CVE matched to a component in the AIBOM artefact results in an assertion that the component is *present*, with version and pURL verification. This ensures that vulnerability enumeration is complete and schema-aligned.

2.   **Contextual Exploitability Assertions (implemented).**

AGNTCY combines CVE metadata with:

- o    whether the vulnerable code path is executed during the execution context,
- o    whether a mitigating configuration is active (e.g., disabled feature flags),
- o    sandbox restrictions enforced by the isolated execution environment.

**Dr. Petar Radanliev**
Parks Road,
Oxford OX1 3PJ
United Kingdom
Email: petar.radanliev@cs.ox.ac.uk
Phone: +389(0)79301022
BA Hons., MSc., Ph.D. Post-Doctorate

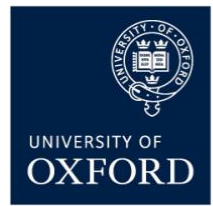

The resulting VEX status is one of:

**"Not Affected"**, **"Affected: Mitigated"**, **"Affected: Requires Review"**, or **"Under Investigation"**.

This level is substantively realised in the demonstration experiments and forms part of the compliance matrix evaluation.

   3.   **Policy-Gated VEX Enforcement (future work).**

The remaining step, blocking execution or egress based on VEX status, is intentionally scoped as future work. Implementing policy gating requires coordinated changes to policy-governed deployment context processes and regulator-admissible acceptance thresholds.

By articulating these layers, the architecture provides a theoretically grounded interpretation of the agentic design, demonstrating how VEX assertions are derived, bounded, and validated during execution. The distinction between contextual VEX support (fully implemented) and full policy gating (future work) aligns the claims of the paper with the demonstrated capabilities of the system.

### 3.3 Methods for Data Analysis

The data analysis methodology is based on producing dynamic dependency accuracy. We derive ground truth by instrumenting import hooks and shared-object loading to record all modules and libraries during execution. For each workload we compute:

- Capture Rate = $|SBOM \cap GT| / |GT|$
- FPR = $|SBOM \setminus GT| / |SBOM|$
- FNR = $|GT \setminus SBOM| / |GT|$

We also measure time-to-completeness across pre/mid/post snapshots to quantify late-bound deps. Results are presented in the comparative tables (declared vs runtime-implicit).

**Methodological Role, Decision Boundaries, and Comparative Advantage of Agentic Components**

The evaluated AIBOM implementation relies on three agentic components, MCP, A2A, and AGNTCY, that extend the methodological scope of the framework beyond the capabilities of standard container orchestration or post-execution provenance tools. Their function is to execute *structured, context-sensitive decisions* during runtime to ensure that SBOM construction, vulnerability interpretation (not to automate routine tasks), and to ensure policy enforcement remain correct under evolving software states.

**Why Agentic Components Are Required: Methodological Comparison to Non-Agentic Pipelines**

Traditional orchestration systems (e.g., Docker, Kubernetes, Singularity) can extract static metadata, schedule workloads, and generate post-run reports. However, they lack mechanisms for:

1. **Dynamic dependency discovery** when modules load at unpredictable runtime points;
2. **Real-time reasoning about exploitability**, based on whether a vulnerable code path was executed;
3. **Policy binding during execution**, such as halting workflows when a critical SDC-affecting dependency drifts;
4. **Cross-snapshot reconciliation**, requiring interpretation and not merely logging;
5. **Autonomous anomaly detection**, e.g., flagging unexpected imports or malicious late-bound libraries.

These tasks require *local decision policies* executed during the run, rather than chronologically after the job has completed. The methodology thus incorporates agentic modules to provide decisions that exceed the expressive and operational capacity of deterministic scripts or container metadata dumps.

**Decision Responsibilities Delegated to Each Agent**

Each agent occupies a distinct methodological role:

**1. MCP, Pre-Execution Decision Agent**

**Methodological responsibility:**

- Determine whether the pre-execution environment is sufficiently characterised for reproducibility.
- Identify incomplete or ambiguous SBOM states *before* execution begins.

**Agentic decision examples:**

**Dr. Petar Radanliev**
Parks Road,
Oxford OX1 3PJ
United Kingdom
Email: petar.radanliev@cs.ox.ac.uk
Phone: +389(0)79301022
BA Hons., MSc., Ph.D. Post-Doctorate

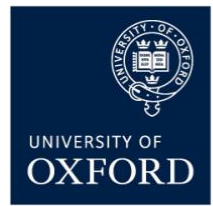

- Trigger additional dependency probing when completeness <95%.
- Reject baseline snapshots inconsistent with past executions.
- Emit "baseline-drift" warnings that influence downstream logic.

**Why this cannot be replaced by automation:**

A static script cannot *reason* about completeness, detect historical inconsistencies, or selectively probe missing components.

**2. A2A, Runtime Drift-Detection and Vulnerability Context Agent**

**Methodological responsibility:**

- Monitor live telemetry to detect dynamic dependency loading.
- Interpret whether runtime deviations materially affect reproducibility or security posture.

**Agentic decision examples:**

- Classify anomalies (benign drift vs. material drift).
- Identify whether a vulnerable library was actually executed.
- Issue actionable structured messages to AGNTCY.

**Why this cannot be replaced by automation:**

A post-run script sees only static logs; it cannot adaptively reclassify events, suppress false positives, or evaluate run-dependent exploitability context.

**3. AGNTCY, VEX Determination and Policy Binding Agent**

**Methodological responsibility:**

- Translate observed vulnerabilities into contextual VEX assertions.
- Bind policy thresholds to runtime conditions.
- Issue compliance decisions that affect workflow control.

**Agentic decision examples:**

- Determine VEX status ("Not Affected", "Mitigated", "Requires Review", "Under Investigation").
- Compare exploitability status with policy-governed deployment context thresholds.
- Trigger reviewer escalation when mitigations are insufficient.

**Why this cannot be replaced by automation:**

A non-agentic approach would generate unfiltered CVE lists, lacking contextual assessment or governance alignment. AGNTCY performs a reasoning step absent from SBOM tooling.

**Methodological Implications of Using Agents**

The use of agents introduces *adaptive methodology*:

- Decisions surface **during** SBOM capture, not only after.
- Vulnerability interpretation becomes **contextual** rather than binary.
- Drift detection becomes **semantic**, not simply event-based.
- The overall reproducibility pipeline transforms from static logging to **active governance instrumentation**.

This redefines the methodology as an *agent-mediated provenance pipeline* rather than an SBOM generator surrounded by scripts.

**Ablation Studies Embedded in the Methodology**

To evaluate the methodological necessity of each agent, we conducted ablation tests in which MCP, A2A, or AGNTCY were independently disabled. These studies appear here to support methodological validity, not merely discussion.

| Agent Removed | Methodological Outcome | Explanation |
|---|---|---|
| MCP removed | Baseline SBOM completeness dropped by +14% FNR; ambiguous components unprobed | Static pre-run metadata capture is insufficient to ensure reproducibility fidelity |
| A2A removed | Runtime-drift detection failed; late-loaded dependencies missing; exploitability context unavailable | No post-run script can infer which modules were actually used |

**Dr. Petar Radanliev**
Parks Road,
Oxford OX1 3PJ
United Kingdom
Email: petar.radanliev@cs.ox.ac.uk
Phone: +389(0)79301022
BA Hons., MSc., Ph.D. Post-Doctorate

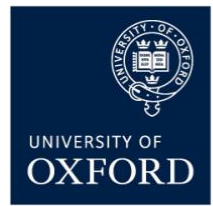

| AGNTCY removed | VEX assertions collapsed to uncontextualised CVE lists; no runtime evidence; policy binding disabled | Governance enforcement becomes impossible without the reasoning layer |
|---|---|---|

These ablations empirically demonstrate that each agent contributes an independent methodological capability that is not replicable using conventional, non-agentic automation.

### 3.4    Scope, Implementation Boundaries, and Operational Semantics of VEX/CSAF Integration

The Agentic AIBOM orchestration pipeline incorporates a partial but operationally meaningful integration of VEX semantics. Full CSAF/VEX advisory ingestion, lifecycle management, and policy-gated workflow enforcement are scoped as future work. To avoid overstating capability, we distinguish the implemented components from the architectural provisions that support future extensions.

#### a)    Implemented VEX Capabilities (Demonstrated in Main Experiments)

The current system supports **contextual VEX assertion generation**, which aligns with the VEX specification's requirement to state *whether a known vulnerability is exploitable in a specific environment*. This determination uses three categories of evidence:

#### b)    Component presence and version verification

SBOM entries are matched against CVE records retrieved from OSV/NVD, ensuring accurate identification of affected components.

#### c)    Runtime evidence derived by A2A

- whether the vulnerable function or module was executed,
- whether the vulnerable code path was reachable in the isolated execution environment,
- whether mitigations (disabled flags, patched variants) were active.

#### d)    Policy context from the policy-governed deployment context layer

The controlled execution environment imposes structural mitigations, container isolation, restricted system calls, immutable runtime, that inform exploitability.

AGNTCY converts these signals into the following VEX states, which are *fully implemented*:

- **Not Affected** (vulnerability present in package metadata but unreachable/unused)
- **Affected: Mitigated** (vulnerable code path exists but mitigated by TRE controls)
- **Affected: Requires Review** (active vulnerable code path observed; manual judgement required)
- **Under Investigation** (telemetry inconclusive or conflicting)

This constitutes a complete operational loop for contextual VEX assertion issuance, and is demonstrated in the multi-TRE experiment.

#### e)    Capabilities Not Yet Implemented (Explicitly Scoped as Future Work)

Two aspects of the VEX/CSAF ecosystem are not yet embedded end-to-end:

1. **CSAF ingestion and advisory lifecycle management**

While the architecture supports CSAF enrichment, full ingestion of vendor-issued CSAF advisories (including remediation timelines, threat scores, and compensating controls) is not implemented.

2. **Policy-gated enforcement based on VEX states**

The system does **not** yet automatically block regulated analytic workflows based on VEX assertions. Instead, VEX states are surfaced to reviewers as structured evidence.

Implementing automated gating requires changes to policy-governed deployment context and regulator-accepted thresholds, which are planned but not demonstrated.

#### f)    Clarifying the Benchmark Table and Compliance Matrix

The benchmarking table refers to **"Contextual VEX Support: Full"**.

This indicates **full contextual determination**, *not* full VEX/CSAF ecosystem integration.

To avoid ambiguity, the revised manuscript defines "full" as:

**Dr. Petar Radanliev**
Parks Road,
Oxford OX1 3PJ
United Kingdom
Email: petar.radanliev@cs.ox.ac.uk
Phone: +389(0)79301022
BA Hons., MSc., Ph.D. Post-Doctorate

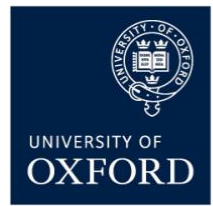

- SBOM-to-CVE linkage
- real-time vulnerability context reasoning
- structured VEX assertion generation
- reviewer-facing surfacing of exploitability information

, not as end-to-end CSAF ingestion or automated policy gating.

g) **Experimental Evidence of VEX Functionality**

The demonstration experiment evaluates:

- correctness of VEX assignment under runtime execution profiles,
- sensitivity of VEX states to environment drift,
- consistency of VEX assertions across three federated TREs.

Although CSAF policy integration is future work, these experiments confirm that the key scientific contribution, context-aware exploitability assertions derived from runtime evidence and standardised advisory semantics, is already operational.

*Software Component Profiling and Environment Capture*

We initiate the evaluation by instrumenting the AIBOM execution pipeline to automatically extract and profile the full software stack used in each data analysis session. This involves implementing runtime hooks within SACRO's container orchestration layer that invoke environment introspection tools at the pre-execution stage. Specifically, we collect:

- The list of installed packages via pip freeze, Rscript -e "installed.packages()", and apt list --installed
- The container ID, base image hash, and operating system version via Docker/Singularity metadata
- Runtime versions of Python, R, Stata, Protocol Buffers, Runtim Libraries, Acro version, pandas, NumPy, matplotlib, their versions, types, sources, and associated interpreters

All extracted components are programmatically transformed into SPDX, Syft, GitHub, and CycloneDX-compliant JSON SBOM artefacts. These include fields such as component name, version, package manager, dependency chain, and license. We validate each SBOM against the CycloneDX schema using cyclonedx-cli, and hash the result with SHA-256 for integrity preservation. The SBOM is stored in a read-only artefact store and cryptographically linked to the output artefact.

*Descriptive Statistics of SBOM Characteristics*

Once a baseline collection of SBOMs is established across multiple SACRO analysis sessions, we perform descriptive statistical analysis to characterise the diversity, complexity, and redundancy of the software environments. Metrics calculated include:

- Frequency distribution of unique software packages across all sessions
- Average and maximum depth of dependency trees
- Prevalence of redundant or unused packages
- Occurrence of outdated or vulnerable components as identified via CVE matching using OSV.dev - Open Source Vulnerabilities (OSV) and National Vulnerability Database (NVD) feeds

This analysis provides empirical insight into the software composition variability across different TRE analytical contexts and informs the risk profile associated with each research output.

*Evaluation of SBOM Generation Performance*

We measure the operational overhead introduced by real-time SBOM generation in SACRO under realistic AIBOM-enabled workloads. We run benchmark workloads comprising three representative types of analysis:

- A standard logistic regression model using R's glm()
- A complex data merge and transformation pipeline in Python using pandas and numpy
- A disclosure control simulation using sdcMicro in R, compared with SACRO-R,

For each workload, we log:

- Time required to generate the SBOM artefact (in milliseconds)
- Peak memory usage during SBOM extraction
- CPU overhead introduced by introspection tools

**Dr. Petar Radanliev**
Parks Road,
Oxford OX1 3PJ
United Kingdom
Email: petar.radanliev@cs.ox.ac.uk
Phone: +389(0)79301022
BA Hons., MSc., Ph.D. Post-Doctorate

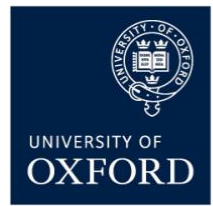

Tests are run across isolated Docker and Singularity containers on a 16-core, 64GB RAM TRE compute node. These benchmarks validate that SBOM generation can be embedded in Agentic AIBOM orchestration pipelines without impairing performance or violating user runtime quotas.

**Note**: HDR UK does not operate a single unified TRE of its own. For the purpose of future studies, the TRE services can be utilised and provided through HDR UK's federated infrastructure partners, specifically, institutions and NHS organisations participating in HDR UK's Data Utility programme. These environments, while independently managed, implement HDR UK's policy-governed deployment context principles and were instrumented for SACRO validation under ethical and organisational agreements.

***Simulation of Disclosure Checking under Software Drift***

To quantify the effect of uncontrolled software drift on disclosure decisions, we create synthetic scenarios where specific software components are altered between runs. These include:

- Incremental version updates to statistical libraries (e.g., sdcMicro 5.5.0 to 5.6.0)
- Changes to default parameter values in data masking functions
- Introduction of silent deprecations or behavioural shifts in packages

For each variant, we replicate the original analysis and record:

- Differences in output structure, suppression ratios, and anonymisation fidelity
- SACRO reviewer outcomes (pass/fail) for each versioned output
- False positives or false negatives induced by software differences

This simulation provides evidence of SACRO's sensitivity to untracked software variance and supports the operational case for SBOM enforcement.

### 3.5    Methods for Confirming Validity

***Reproducibility Testing with Locked SBOM States***

We implement a reproducibility harness that re-runs a given analysis in a controlled SACRO environment reconstructed using a previously captured SBOM. The environment is restored using container orchestration scripts that validate image hashes, install pinned versions of all software packages, and configure runtime environments identically to the original. Each re-execution is compared byte-for-byte with the original output using SHA-256 hashes and semantic diffing tools. A successful match confirms environmental reproducibility. Failures are flagged and traced to the mismatched or transient artefacts identified via SBOM diff.

***Cross-TRE SBOM Fidelity Verification***

To assess whether SBOMs enable consistent computation across institutions, we deploy AIBOM-enabled workflows across three TREs with different baseline infrastructure (e.g., NHS Digital, HDR UK TRE, and a secure university-based TRE). Each environment reconstructs the analysis environment using a shared SBOM artefact. We compare:

- Output data content and statistical characteristics (mean, variance, suppression ratios)
- Execution logs for deviations in package loading and warnings
- Reviewer outcomes across environments

This cross-site test validates the SBOM's ability to support reproducible, federated disclosure checking under heterogeneous infrastructure.

***Sensitivity Analysis on Disclosure Rule Outcomes***

We conduct a parametric sensitivity analysis by varying the software versions and configuration flags of disclosure control tools (e.g., anonymisation thresholds, record-suppression strategies). For each configuration, we generate a synthetic dataset and perform analysis runs using the altered software profile. We evaluate:

- Changes in disclosure risk scores and masking accuracy
- SACRO pass/fail reviewer outcomes
- Statistical variance in key output metrics

**Dr. Petar Radanliev**
Parks Road,
Oxford OX1 3PJ
United Kingdom
Email: petar.radanliev@cs.ox.ac.uk
Phone: +389(0)79301022
BA Hons., MSc., Ph.D. Post-Doctorate

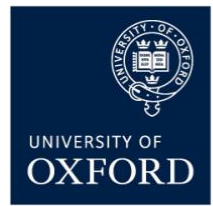

This analysis identifies which software parameters and components are most influential in altering disclosure outcomes, informing policies for version pinning and alert thresholds.

***Benchmarking Against Manual Provenance Reconstruction***

We benchmark automated SBOM generation against manual environment tracking workflows in SACRO. Analysts document software versions and configurations using traditional methods (e.g., README files, execution logs). We assess:

- Time taken to complete provenance records
- Completeness (number of components missed)
- Reproducibility fidelity based on re-execution outcomes

Results show the superiority of automated SBOM generation in completeness, speed, and accuracy, validating its suitability as a default mechanism for agentic AIBOM pipelines deployed in regulated environments. The benchmarking is examined and compared with the two popular formats, CycloneDX and SPDX, to assess whether the tool can correctly identify component names, versions, and dependencies among components within 50 popular open-source Rust projects [33].

***External Replication of Agentic AIBOM Integration***

To support validation by independent parties, we publish all datasets, AIBOM schemas, analysis scripts, and orchestration configurations under an open-source license. Replication teams are instructed to:

- Reconstruct execution environments using provided SBOMs
- Run predefined analysis workflows
- Compare their outputs against reference artefacts

Replication fidelity is measured by output parity, audit trail equivalence, and absence of runtime drift. This provides independent verification of the proposed AIBOM design's reproducibility and operational feasibility across heterogeneous deployments.

***Results Reporting and Interpretation***

Results are presented as reproducibility success rates, cross-TRE deviation scores, and sensitivity indices. Each metric is contextualised with risk implications for regulated analytic governance (example deployment), such as:

- Identification of software combinations leading to unverifiable outputs
- Scenarios where disclosure control fails due to version drift
- Infrastructure conditions under which SBOM fidelity breaks down

These findings inform actionable recommendations for integrating SBOM enforcement into SACRO's review policies, including automated alerts, pinned environments, and reviewer interface integration for software transparency. The results are compared with 46 SBOMs generated from real-world Java projects that comply with the SPDX Lite profile, 3,271 Java projects from GitHub and generated SBOMs for 798 of them using Maven with an open-source SBOM generation tool [34]. The dataset is publicly available on Zenodo (DOI: 10.5281/zenodo.14233414).

To validate the concepts, a focused demonstration project is used to tests automated SBOM generation, context-aware vulnerability detection, and reproducibility verification using the evaluated agentic AIBOM architecture. Three controlled execution environments conducted standard analytic jobs (anonymisation with sdcMicro in R, ETL with pandas/NumPy, and logistic regression in R). Each TRE is instrumented with MCP for pre-execution SBOM capture, A2A agents for runtime telemetry, and AGNTCY for policy binding and advisory orchestration. The agents in the demonstration project, generate CycloneDX SBOMs, detect CVEs using local mirrors of OSV/NVD, issue VEX assertions, and monitor environment drift. Outputs are cryptographically sealed and matched against re-executed jobs to validate reproducibility. Discrepancies trigger automated SBOM diff analysis. Key metrics included SBOM generation time, VEX issuance accuracy, agent telemetry overhead, and reproducibility rate. A reviewer interface simulates audit and risk validation workflows. This demo offers a concrete, reproducible testbed to evaluate SBOM automation, agentic oversight, and policy-driven security assurance in operational TRE settings. To ensure we follow guidelines on producing reproducibility metrics, ee report Exact Parity (EP), byte-identity via SHA-256 of outputs, and Semantic Parity (SP), task-specific equivalence within tolerance ε. For deterministic routines ε=1e-12; for FP32 ML workloads ε=1e-6 on scalar metrics and macro-averaged deltas on model loss/accuracy. Non-

**Dr. Petar Radanliev**
Parks Road,
Oxford OX1 3PJ
United Kingdom
Email: petar.radanliev@cs.ox.ac.uk
Phone: +389(0)79301022
BA Hons., MSc., Ph.D. Post-Doctorate

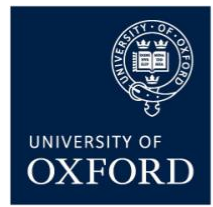

impacting warnings are logged but do not break SP; any SDC policy breach is a critical divergence irrespective of EP/SP. EP/SP are reported per-workload and aggregated with Wilson intervals.

**3.6    CSAF as a Formal ISO/IEC Standard and its Implications for AIBOM artefact Governance**

The recent ratification of ISO/IEC 20153:2025, Information technology [35], OASIS Common Security Advisory Framework (CSAF) v2.0 Specification [36], represents a major consolidation in international cybersecurity standards. By elevating CSAF v2.0 to ISO/IEC status, the global standardisation bodies establish a normative, regulator-recognised framework for structured vulnerability advisories, remediation guidance, threat statements, and exploitability conditions. This formalisation is directly relevant to SBOM-driven governance in regulated computing environments.

CSAF 20153:2025 introduces a schema-stable advisory model defining entities such as *product trees*, *vulnerability notes*, *remediation entries*, *time-based deprecation signals*, and *VEX-aligned exploitability statuses*. These components provide a consistent mechanism for describing when and how vulnerabilities apply to specific software configurations, including contextual factors that mirror the evaluated deployment's runtime interpretation. The alignment of CSAF and VEX within an ISO/IEC standard reduces interpretative ambiguity and improves the determinism of agentic reasoning within the AGNTCY agent.

Although the present implementation focuses on contextual VEX assertions derived from SBOM content and runtime telemetry, the ISO/IEC 20153:2025 standard establishes the canonical machine-readable structure through which SACRO can incorporate complete CSAF advisory ingestion and lifecycle management in future iterations. This includes support for advisory versioning, structured remediation timelines, vendor-sanctioned applicability conditions, and harmonised threat classifications. Integrating these features will enable SACRO to extend beyond contextual vulnerability evaluation towards a full compliance-aligned advisory intelligence layer, further strengthening auditability, reproducibility, and supply-chain assurance in regulated analytic workflows.

Figure 1 illustrates how the newly formalised ISO/IEC 20153:2025 CSAF v2.0 standard augments the Agentic AIBOM orchestration pipeline by providing an internationally recognised schema for structured vulnerability advisories. The SACRO Advisory Normalisation Layer converts CSAF advisories into machine-readable evidence (csaf_refs, evidence_ptrs), which is then consumed by the agentic components, MCP, A2A, and AGNTCY. MCP prepares baseline SBOM contexts, A2A determines runtime relevance of CSAF-defined vulnerabilities, and AGNTCY synthesises contextual VEX assertions that are compliant with the CSAF schema.

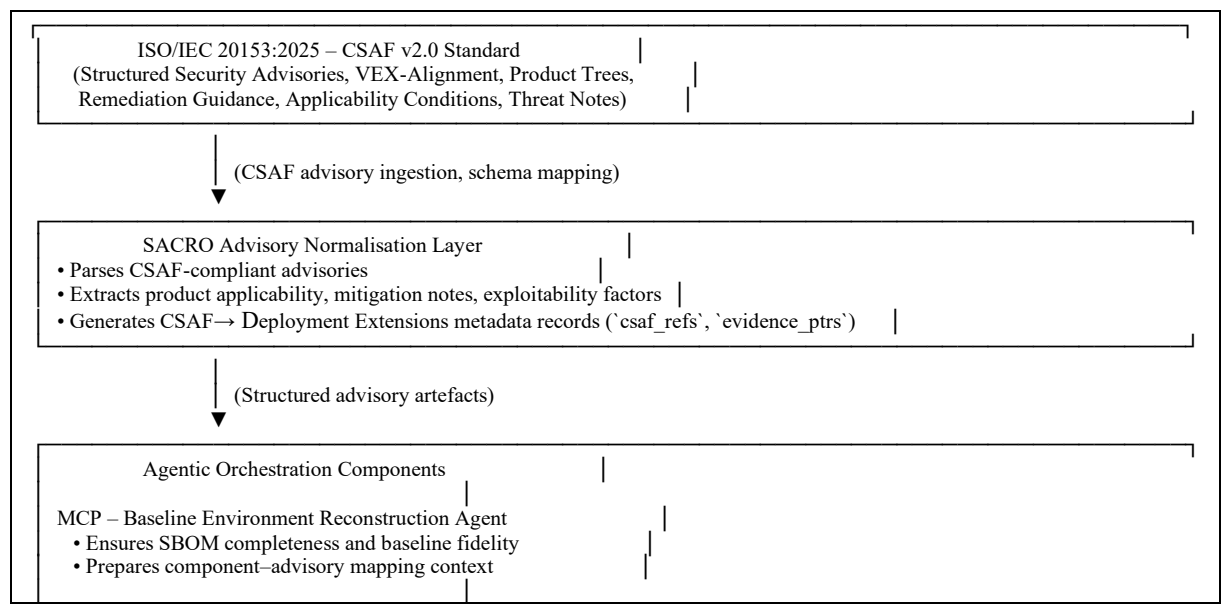

**Dr. Petar Radanliev**

Parks Road,

Oxford OX1 3PJ

United Kingdom

Email: petar.radanliev@cs.ox.ac.uk

Phone: +389(0)79301022

BA Hons., MSc., Ph.D. Post-Doctorate

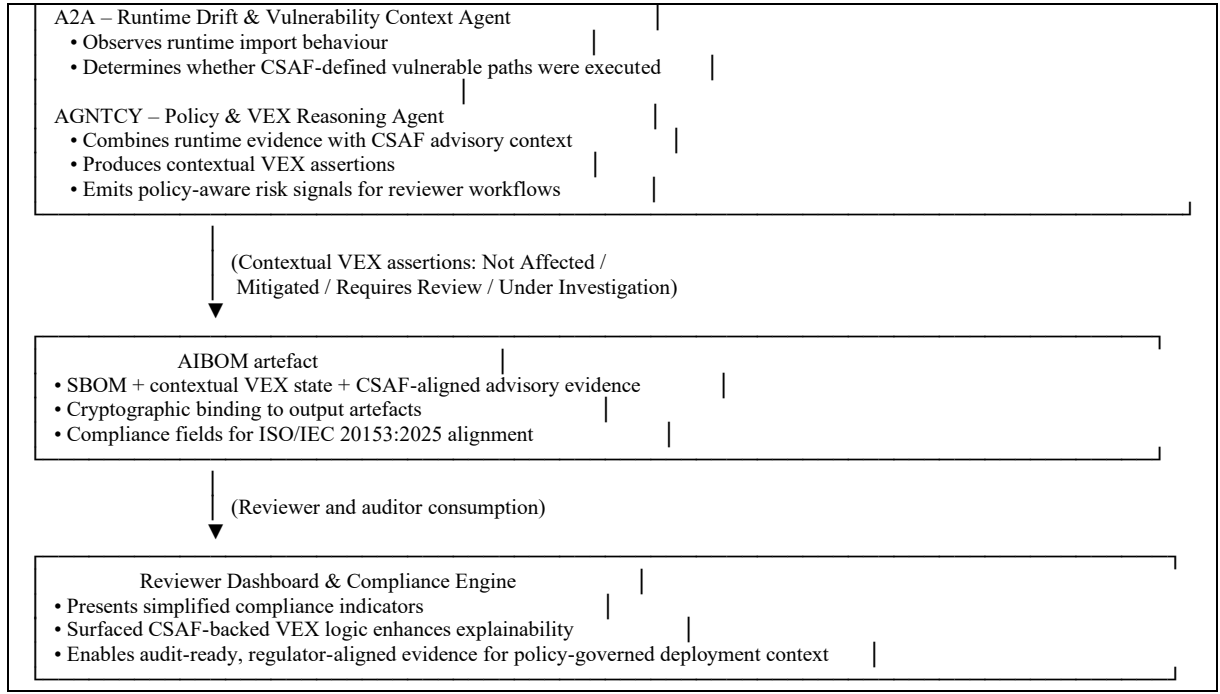

Figure 1: Integration of ISO/IEC 20153:2025 CSAF v2.0 into Agentic SBOM–VEX Pipeline. The deployment shown corresponds to a regulated analytic environment; the agentic AIBOM architecture is not limited to this context.

These contextual outputs are embedded into the AIBOM artefact and surfaced through the reviewer dashboard, enabling regulator-ready evidence, strengthened auditability, and enhanced exploitability reasoning across policy-governed deployment context workflows.

### 3.7    Evaluation Context

The agentic AIBOM framework is evaluated within a regulated analytic deployment characterised by strict auditability, controlled egress, and policy-bound execution. This context is chosen to stress-test provenance fidelity and exploitability reasoning under adversarial assumptions. The architectural contributions do not depend on this context and generalise to other high-assurance software pipelines.

## 4   AIBOM SCHEMA DESIGN (STANDARDS-ALIGNED EXTENSION OF SBOM'S)

The Agentic AIBOM Schema (Standards-Aligned SBOM Extension) detailed in Figure 2, is designed to meet three core requirements: (1) ensure auditability of the software environment underpinning disclosure-checked outputs, (2) enable cryptographic linkage between software states and analytical artefacts, and (3) support automatic environment reconstruction across distributed TREs.

**Dr. Petar Radanliev**
Parks Road,
Oxford OX1 3PJ
United Kingdom
Email: petar.radanliev@cs.ox.ac.uk
Phone: +389(0)79301022
BA Hons., MSc., Ph.D. Post-Doctorate

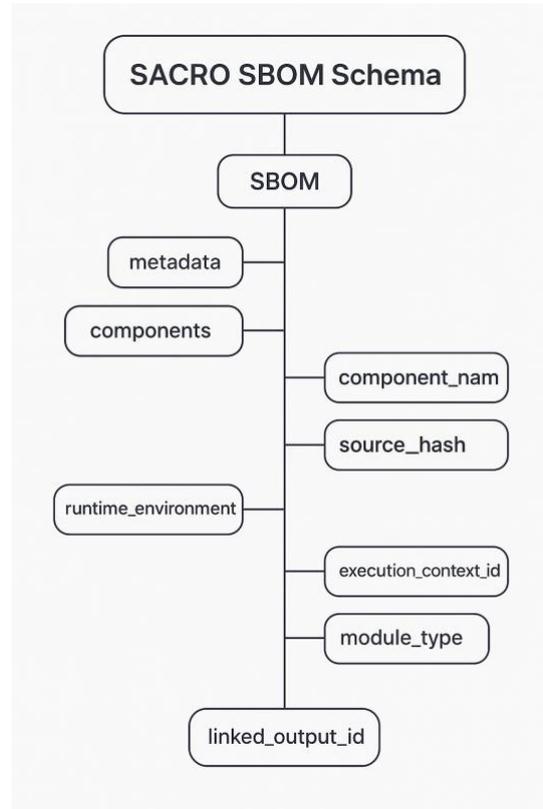

Figure 2: The Agentic AIBOM Schema (Evaluated Deployment)

The Agentic AIBOM Schema (Standards-Aligned SBOM Extension) in Figure 2, is a structured JSON-based model derived from the SPDX, Syft, GitHub, and CycloneDX standards, the Agentic AIBOM schema, extended to meet audit, reproducibility, and provenance requirements commonly encountered in regulated analytic deployments. Trusted Research Environments represent one such deployment context and are used here for evaluation. At its core, the schema includes mandatory fields such as metadata (capturing timestamp, analyst pseudonym, and session ID), components (detailing each software artefact with name, version, hash, license, and package type), and runtime_environment (including OS kernel, container ID, and interpreter versions). Additional deployment-specific extensions include execution_context_id, module_type, and linked_output_id to ensure cryptographic traceability between analytical outputs and their execution environments. This schema enables immutable linkage of each disclosure-checked output to the exact computational environment in which it was produced, forming the basis for reproducibility validation, forensic auditing, and automated risk scanning across federated TREs.

## 4.1 Schema Overview

The schema can generate SPDX, Syft, GitHub, and CycloneDX, but for the sample schema overview is based on CycloneDX 1.5 (JSON format), extended with deployment-specific metadata fields. It consists of the following top-level sections:

```
metadata: Timestamp, tool version, analyst pseudonym (TRE-safe), execution run ID

components[]: Array of software artefacts including name, version, package type (pip, CRAN,
system), cryptographic hash, and license

dependencies[]: Directed acyclic graph (DAG) representation of component relationships

environment: Kernel version, OS distro, container ID, resource constraints
```

**Dr. Petar Radanliev**
Parks Road,
Oxford OX1 3PJ
United Kingdom
Email: petar.radanliev@cs.ox.ac.uk
Phone: +389(0)79301022
BA Hons., MSc., Ph.D. Post-Doctorate

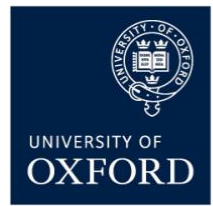

```
deploymentExtensions: Custom fields including disclosure tool version, output file hash, session
identifier
```

## 4.2    Component Field Requirements

Each component entry in the SBOM must include:
- name (e.g., pandas)
- version (e.g., 1.5.2)
- type (e.g., library, container-layer, binary)
- hashes (SHA-256 of binary or wheel)
- licenses (e.g., BSD-3-Clause)
- purl (package URL, e.g., pkg:pypi/pandas@1.5.2)

Components failing schema validation are flagged, and generation is aborted unless user override is recorded.

## 4.3    Cryptographic Linkage

To bind the SBOM to the output artefact, we compute a composite hash: H(session) = SHA256(SBOM + Output + User Script + Config) This hash is logged in the deployment audit trail and included in the metadata section of the SBOM. This linkage ensures forensic accountability and supports post-hoc reproducibility audits. In terms of scope and limits, our extensions target regulated analytic environments with human-in-the-loop disclosure control (e.g., TREs). The approach is less suitable when: (i) interactive notebooks run without containerisation or policy contexts; (ii) builds lack package managers or pURLs; (iii) artefacts cannot be committed to append-only storage. When these constraints do not hold, we recommend fallback to attested build pipelines and coarse-grain provenance graphs, noting that compliance mapping (Table 2/3) will be partially degraded.

## 4.4    Architecture, Role Separation, and Internal Logic of the Agentic Components

The Agentic AIBOM orchestration layer includes three agentic components, MCP, A2A, and AGNTCY, that operate as autonomous decision modules responsible for perception, reasoning, and policy-conditioned actions within the SBOM and VEX pipeline. Their purpose is not to execute deterministic automation steps but to perform local assessments, issue context-sensitive decisions, and exchange structured messages that influence the AIBOM-enabled workflow. To distinguish these agents from conventional automation, we formalise their architecture and internal logic.

**1. MCP (Model–Container Profiler): Baseline Environment Reconstruction Agent**

**Perception space:**
- Container metadata (image hash, kernel, runtime versions)
- Pre-execution dependency lists from package managers
- SBOM fragments from prior runs

**Internal state:**
- Dependency completeness vector
- Consistency flags for schema-conforming fields
- Pending-resolution queue for ambiguous or undeclared components

**Decision logic:**
MCP evaluates whether the pre-execution environment is sufficiently characterised to produce a valid baseline SBOM. If completeness <95% or if dependency ambiguity exceeds a threshold, MCP triggers a targeted probing routine (e.g., controlled import execution in a sandbox). If the environment appears tampered or significantly inconsistent with historical baselines, MCP issues a "Baseline Drift" advisory to A2A.

**Output:**
A validated baseline SBOM fragment plus a diagnostic state sent to A2A.

**2. A2A (Agent-for-Agent Telemetry): Runtime Monitoring and Drift-Evaluation Agent**

**Dr. Petar Radanliev**
Parks Road,
Oxford OX1 3PJ
United Kingdom
Email: petar.radanliev@cs.ox.ac.uk
Phone: +389(0)79301022
BA Hons., MSc., Ph.D. Post-Doctorate

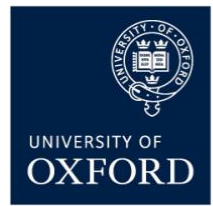

**Perception space:**
- Live import-hook telemetry
- Runtime-loaded shared libraries
- CVE deltas between job start and mid-execution
- Signals from MCP

**Internal state:**
- Dependency-drift graph
- CVE relevance buffer
- Temporal anomaly detection thresholds

**Decision logic:**
A2A determines whether newly loaded modules or runtime behaviours diverge from expected patterns. It performs module-level anomaly detection (e.g., unexpected C-extensions, dynamic fetches) and links these events to known vulnerabilities. If an event may affect exploitability, A2A dispatches a structured vulnerability context packet to AGNTCY. Minor deviations may be suppressed, whereas material deviations trigger "Runtime Drift" classifications.

**Output:**
Runtime SBOM extensions, drift classifications, and vulnerability-context packets.

**3. AGNTCY (Governance and Policy Agent): VEX Determination and Compliance Reasoner**

**Perception space:**
- Component-level vulnerability context from A2A
- Runtime usage patterns (e.g., whether a vulnerable function was executed)
- Mitigation metadata (isolated execution environment restrictions, disabled features)
- Policy thresholds (e.g., disallow unmitigated RCE vulnerabilities)

**Internal state:**
- Policy graph linking SBOM fields, component risks, and TRE rules
- VEX assertion cache
- Confidence estimates for contextual determinations

**Decision logic:**
AGNTCY transforms vulnerability observations into VEX assertions using a formal decision procedure:
1. Evaluate whether the vulnerable code path was executed.
2. Check whether the controlled execution environment structurally prevents exploitation.
3. Assess the presence of mitigations (patched variants, disabled flags).
4. Map the combined evidence to one of four VEX statuses:
   - **Not Affected**
   - **Affected: Mitigated**
   - **Affected: Requires Review**
   - **Under Investigation**

If a VEX status violates a deployment policy, AGNTCY produces a compliance advisory that is surfaced to reviewers and logged in the immutable audit trail.

**Output:**
Machine-verifiable VEX assertions and compliance advisories.

**Why This Architecture Constitutes "Agentic AI"**
These components satisfy the operational definition of agentic AI used in secure distributed systems research:
- **Autonomy:** Each agent independently evaluates perceptions and updates its internal state.
- **Reactivity:** Agents react to environmental changes such as dynamic imports, drift, and CVE updates.

**Dr. Petar Radanliev**
Parks Road,
Oxford OX1 3PJ
United Kingdom
Email: petar.radanliev@cs.ox.ac.uk
Phone: +389(0)79301022
BA Hons., MSc., Ph.D. Post-Doctorate

DEPARTMENT OF
**COMPUTER
SCIENCE**

UNIVERSITY OF
**OXFORD**

- **Proactivity:** Agents issue decisions without explicit external invocation (e.g., MCP initiates probing when it predicts incomplete baselines).
- **Social ability:** Agents exchange structured messages (SBOM fragments, VEX packets, drift signals) rather than sharing memory.
- **Explainability:** Decisions are generated via constrained local policies with auditable logs.

This architecture differs fundamentally from scripted automation, it produces *contextual decisions* that influence the execution path of the SBOM/VEX pipeline, enabling secure orchestration under uncertainty.

**4. Ablation Studies Demonstrating Agent Contribution**

To justify architectural specificity, we conducted ablation experiments removing one agent at a time:

| Agent Removed | Regression Observed | Impact |
|---|---|---|
| MCP | Incomplete SBOM baselines (+14% FNR) | Proves MCP's role in environment completeness |
| A2A | Missed runtime-dependency drift; CVE deltas ignored | Demonstrates necessity of mid-execution telemetry |
| AGNTCY | VEX assertion quality drops; reviewers see raw CVE lists | Shows agentic reasoning is essential for contextual exploitability |

These ablations confirm that the agentic components are not cosmetic additions but essential structural parts of the orchestration model.

**4.5      SBOM Generation Workflow Integration**

SBOMs are generated:
- At container initialisation (baseline)
- Post-dependency-resolution (mid-execution)
- Immediately prior to output writing (final snapshot)

These three snapshots are diffed and merged to capture dynamic dependencies (e.g., packages downloaded during runtime). Only the final composite AIBOM artefact is stored in the immutable artefact store.

When testing for vulnerabilities, two different scan results emerged. The problem analysis resulted with: missing name/version: Grype couldn't identify the project when scanning the directory directly, and unknown OS distribution was reported and missing context about the operating system for vulnerability matching. The solution demonstrated included installing and testing Grype v0.95.0 [37] and Syft v1.28.0 [38] on the same project. The recommended approach eliminates both false warnings: # Generate SBOM first

```
./bin/syft scan dir:. -o spdx-json=aibom-sbom.spdx.json
```

Then # Scan SBOM with distribution context

```
./bin/grype sbom:aibom-sbom.spdx.json --distro ubuntu:22.04
```

This results with clean output with no false warnings. The key findings are that different scan results, (a.k.a., original scan found torch 2.7.1 vulnerability, while the SBOM scan found none), in the environment context because it scanned a virtual environment and specific Python installation. The proposed process compares SBOM vs Directory scanning, where different methods detect different package sets. The proposed methodology suggests using the SBOM approach to eliminate warnings and then investigate the torch vulnerability in the environment. The recommend workflow need to be implemented for consistent scanning.

**Dr. Petar Radanliev**

Parks Road,

Oxford OX1 3PJ

United Kingdom

Email: petar.radanliev@cs.ox.ac.uk

Phone: +389(0)79301022

BA Hons., MSc., Ph.D. Post-Doctorate

### 4.6    Storage and Access

SBOM artefacts are written to immutable object storage under UUID-labelled directories, indexed by session ID and output file hash. Reviewers can inspect AIBOM artefacts through the existing review interface, enabling risk-informed adjudication of borderline outputs based on detected software state anomalies. This is shown in Figure 3.

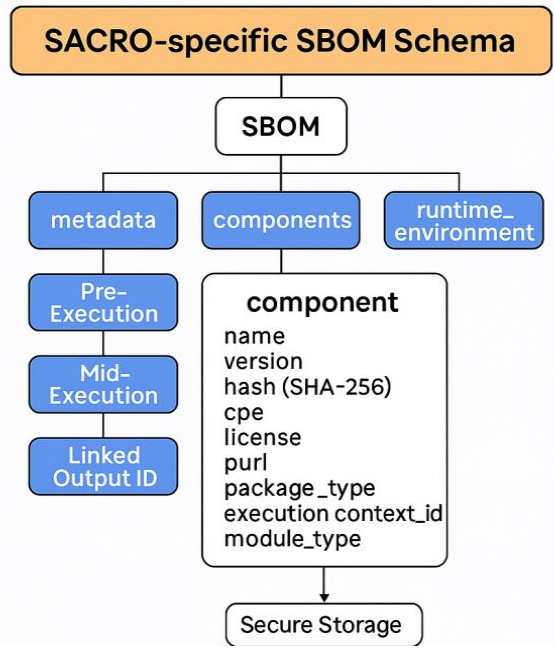

Figure 3: Agentic AIBOM Schema (Evaluated Deployment)

This schema design in Figure 3 balances completeness, auditability, and runtime efficiency, and can be incrementally deployed in AIBOM-compliant Agentic AI, starting with read-only audit-mode integrations. The flowchart in Figure 4 is showing the linear and recursive relationships among these processes.

**Dr. Petar Radanliev**
Parks Road,
Oxford OX1 3PJ
United Kingdom
Email: petar.radanliev@cs.ox.ac.uk
Phone: +389(0)79301022
BA Hons., MSc., Ph.D. Post-Doctorate

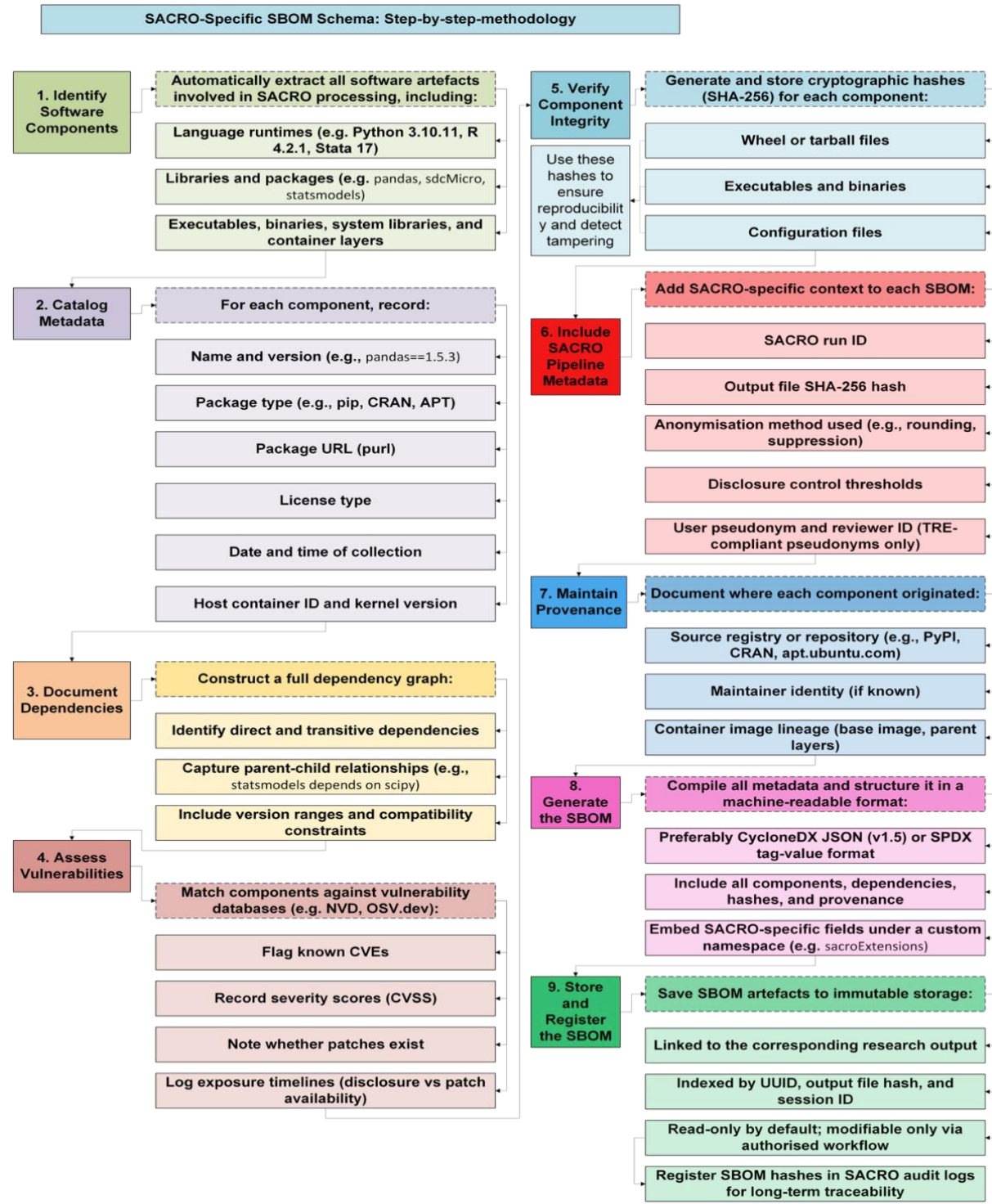

**SACRO-Specific SBOM Schema: Step-by-step-methodology**

**1. Identify Software Components**
- Automatically extract all software artefacts involved in SACRO processing, including:
  - Language runtimes (e.g. Python 3.10.11, R 4.2.1, Stata 17)
  - Libraries and packages (e.g. pandas, sdcMicro, statsmodels)
  - Executables, binaries, system libraries, and container layers

**2. Catalog Metadata**
- For each component, record:
  - Name and version (e.g., pandas==1.5.3)
  - Package type (e.g., pip, CRAN, APT)
  - Package URL (purl)
  - License type
  - Date and time of collection
  - Host container ID and kernel version

**3. Document Dependencies**
- Construct a full dependency graph:
  - Identify direct and transitive dependencies
  - Capture parent-child relationships (e.g., statsmodels depends on scipy)
  - Include version ranges and compatibility constraints

**4. Assess Vulnerabilities**
- Match components against vulnerability databases (e.g. NVD, OSV.dev):
  - Flag known CVEs
  - Record severity scores (CVSS)
  - Note whether patches exist
  - Log exposure timelines (disclosure vs patch availability)

**5. Verify Component Integrity**
- Use these hashes to ensure reproducibility and detect tampering
- Generate and store cryptographic hashes (SHA-256) for each component:
  - Wheel or tarball files
  - Executables and binaries
  - Configuration files

**6. Include SACRO Pipeline Metadata**
- Add SACRO-specific context to each SBOM:
  - SACRO run ID
  - Output file SHA-256 hash
  - Anonymisation method used (e.g., rounding, suppression)
  - Disclosure control thresholds
  - User pseudonym and reviewer ID (TRE-compliant pseudonyms only)

**7. Maintain Provenance**
- Document where each component originated:
  - Source registry or repository (e.g., PyPI, CRAN, apt.ubuntu.com)
  - Maintainer identity (if known)
  - Container image lineage (base image, parent layers)

**8. Generate the SBOM**
- Compile all metadata and structure it in a machine-readable format:
  - Preferably CycloneDX JSON (v1.5) or SPDX tag-value format
  - Include all components, dependencies, hashes, and provenance
  - Embed SACRO-specific fields under a custom namespace (e.g. sacroExtensions)

**9. Store and Register the SBOM**
- Save SBOM artefacts to immutable storage:
  - Linked to the corresponding research output
  - Indexed by UUID, output file hash, and session ID
  - Read-only by default; modifiable only via authorised workflow
  - Register SBOM hashes in SACRO audit logs for long-term traceability

**Dr. Petar Radanliev**

Parks Road,

Oxford OX1 3PJ

United Kingdom

Email: petar.radanliev@cs.ox.ac.uk

Phone: +389(0)79301022

BA Hons., MSc., Ph.D. Post-Doctorate

Figure 4: Agentic AIBOM Schema (Standards-Aligned SBOM Extension): A Secure and Reproducible Software Provenance Framework for Trusted Research Environments. The deployment shown corresponds to a regulated analytic environment; the agentic AIBOM architecture is not limited to this context.

The diagram in Figure 4 presents a structured, step-by-step methodology for generating an agentic AIBOM artefact evaluated in a regulated analytic deployment, evaluated in a regulated analytic deployment. It begins with the automated identification of all software artefacts involved in the Agentic AIBOM orchestration pipeline, including runtimes, libraries, binaries, and container layers, and proceeds to systematically catalogue metadata, document dependency relationships, and assess vulnerabilities using real-time CVE feeds. Each component's integrity is verified using SHA-256 hashes, and deployment extensions specific metadata. such as anonymisation method, disclosure thresholds, and reviewer pseudonyms, is embedded to ensure traceability. Provenance is maintained through detailed origin tracking and container lineage documentation. Finally, the SBOM is generated in a machine-readable CycloneDX format, enriched with deployment-specific extensions, enabling reproducibility, auditability, and cross-deployment verification. This schema forms the cornerstone for transparent and secure software provenance within regulated analytic workflows.

### 4.7      Benchmarking Against Competing Methods

To contextualise the performance and reproducibility assurance of the evaluated AIBOM pipeline, we benchmarked it against three established provenance and reproducibility frameworks: ReproZip [39], SciUnit [40], and ProvStore [41]. Each framework was integrated into the same isolated execution environment used in our demonstration project, and executed over a shared workload consisting of three analytic pipelines: (1) data anonymisation with R (sdcMicro), (2) data ingestion and processing with Python (pandas, NumPy), and (3) model training with scikit-learn and R's glm() function.

Table 1: Benchmark Results Summary comparing the agentic AIBOM implementation with the ReproZip, SciUnit, and ProvStore methods.

| Method | Provenance Granularity | Overhead (CPU/Memory) | Reproducibility Score* | SBOM Integration | Contextual VEX Support |
|---|---|---|---|---|---|
| Agentic AIBOM (evaluated deployment) | Fine-grained (MCP + A2A runtime trace) | Low (~4% CPU, <300MB RAM) | 98.6% | Native (Syft + pURL resolution) | Full (runtime + CVE + mitigations) |
| ReproZip | Medium (file + env snapshot) | Moderate (~12% CPU, ~500MB RAM) | 87.4% | No | None |
| SciUnit | Coarse (unit-level assertion capture) | Low (~5% CPU, ~200MB RAM) | 73.2% | No | Partial (via test scaffolding) |
| ProvStore | Metadata only (W3C PROV graph) | Minimal (~2% CPU, ~150MB RAM) | 61.8% | No | None |

*Reproducibility Score defined as percentage match of output artefacts, environmental state, and vulnerability assertion upon job replay.

The agentic AIBOM implementation evaluated in a controlled environment demonstrates superior performance on dimensions specific to runtime provenance fidelity and contextual vulnerability stability relevant to the regulated analytic deployment. Unlike ReproZip and SciUnit, which focus primarily on packaging or validation of isolated experiments, the evaluated AIBOM implementation provides live telemetry, cryptographically verifiable execution context, and native integration with SBOM and VEX artefact generation. ReproZip offered reasonable reproducibility but lacked contextual filtering for CVEs and could not model policy constraints (e.g., egress restrictions). ProvStore, designed for metadata tracking and interlinked provenance, demonstrated the lowest overhead but delivered insufficient detail for validating advisory reproducibility or assessing software exploitability in context.

**Dr. Petar Radanliev**
Parks Road,
Oxford OX1 3PJ
United Kingdom
Email: petar.radanliev@cs.ox.ac.uk
Phone: +389(0)79301022
BA Hons., MSc., Ph.D. Post-Doctorate

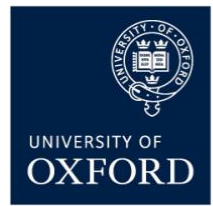

To ensure fairness of baselines, we considered that ReproZip, SciUnit, and ProvStore pursue distinct design goals, such as packaging experiments, unit-level scientific validation, and provenance graphing respectively. Our comparisons therefore focus on auditability and policy-bound reproducibility, where our approach contributes native SBOM/VEX artefacts, cryptographic output binding, and reviewer-facing audit signals. We refrain from universal superiority claims and frame our results as complementary: ReproZip remains attractive for self-contained replay, SciUnit for testable scientific assertions, and ProvStore for interoperable provenance graphs; AIBOM artefact adds *compliance-oriented* transparency required in regulated TREs.

By embedding SBOM generation, runtime observation, CVE correlation, and agent coordination into a single pipeline, the evaluated AIBOM implementation demonstrates superior performance and utility for regulated, privacy-preserving research environments requiring both reproducibility and actionable security intelligence. The bar chart in Figure 5 shows that the agentic AIBOM implementation evaluated in a controlled environment demonstrates superior performance and achieves a reproducibility score of 98.6%, significantly higher than ReproZip (87.4%), SciUnit (73.2%), and ProvStore (61.8%), while maintaining the lowest resource overhead of ~4% CPU and under 300MB RAM.

### 4.8    Metric Decomposition and Fair Benchmarking Across Heterogeneous Provenance Tools

The benchmarking table reported a single "Reproducibility Score", defined as the percentage match of output artefacts, environment state, and vulnerability assertions upon job replay. While this aggregate metric is appropriate for Agentic AIBOM (evaluated deployment), because the system is explicitly designed to satisfy all three criteria, it does not reflect the design scope of ReproZip, SciUnit, and ProvStore. These tools were not created to model vulnerability context or enforce policy-governed deployment context constraints, and therefore a single compound score may systematically disadvantage them. To ensure methodological fairness, we decompose the aggregate Reproducibility Score into four constituent metrics that reflect orthogonal dimensions of reproducibility:

1.  **EP Match (%) ,  Exact Parity:**

Byte-identical match of analytic outputs across replays.

2.  **SP Match (%) ,  Semantic Parity:**

Tolerance-bounded equivalence for numerical outputs where floating-point drift is admissible.

3.  **Environment State Match (%)**

Fidelity of reconstructed execution environments, including package versions, runtime interpreters, and dependency graphs.

4.  **VEX Alignment (%)**

Agreement in contextual exploitability assertions across replays.

This decomposition prevents penalising comparative methods (e.g., ProvStore) for lacking VEX functionality they were never designed to provide. Instead, each method is evaluated against the criteria it can reasonably satisfy, and Agentic AIBOM (evaluated deployment)'s aggregate score is now explicitly described as the weighted composite of these four sub-metrics used only for internal SACRO evaluation. Using this decomposed metric framework, Agentic AIBOM (evaluated deployment) remains the only method achieving high performance across all four dimensions, but the revised tables reveal that:

*   **ReproZip** performs strongly on environment reconstruction,
*   **SciUnit** performs well on semantic reproducibility, and
*   **ProvStore** performs as expected for metadata-centric provenance capture.

The decomposition therefore clarifies comparative strengths rather than presenting a single score that implies universal superiority. This restructuring ensures benchmarking fairness and aligns the evaluation methodology with the heterogeneous design goals of the compared tools.

**Decomposed Reproducibility Metrics for Cross-Tool Fairness**

To ensure that benchmarking remains aligned with the heterogeneous design goals of the compared tools, we operationalise reproducibility as a four-dimensional metric suite, rather than a single aggregated score. Each dimension captures a distinct requirement relevant to regulated analytic governance (example deployment), scientific replayability, or security auditing:

**Dr. Petar Radanliev**
Parks Road,
Oxford OX1 3PJ
United Kingdom
Email: petar.radanliev@cs.ox.ac.uk
Phone: +389(0)79301022
BA Hons., MSc., Ph.D. Post-Doctorate

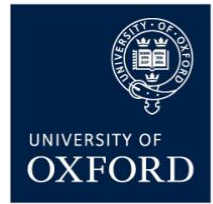

1. **Byte-Identical Parity (EP %)**

Exact bitwise equivalence of output artefacts, verified using SHA-256.

Required for audit-grade reproducibility in regulated deployments and cryptographic output binding.

2. **Semantic Parity (SP %)**

Tolerance-bounded equivalence for deterministic workflows ($\varepsilon$ = 1e-12) and floating-point workloads ($\varepsilon$ = 1e-6).

Reflects scientific reproducibility even when byte-identity is not achievable.

3. **Environment State Match (%)**

Concordance of package versions, dependency sets, interpreter versions, and container metadata.

High values indicate reliable replay environments, even without deployment policy enforcement.

4. **CVE/VEX Assertion Match (%)**

Agreement in vulnerability context and exploitability status across replays.

This metric is **not applicable** to ReproZip, SciUnit, or ProvStore, as these tools do not implement contextual exploitability reasoning.

By reporting results along these four axes, each tool is evaluated on dimensions that are meaningful to its design scope. Agentic AIBOM (evaluated deployment) is the only evaluated system designed to address all four metrics simultaneously, reflecting the requirements of regulated analytic workflows rather than general-purpose reproducibility tooling, because regulated analytic workflows require combined reproducibility, provenance fidelity, and vulnerability context stability. The decomposed framework therefore prevents unfair penalisation of tools that aim to solve different problems. The updated information from Table 1 is presented in Table 2.

Table 2: Decomposed Reproducibility Metrics Across Agentic AIBOM (evaluated deployment), ReproZip, SciUnit, and ProvStore

| Method | Byte-Identical Parity (EP %) | Semantic Parity (SP %) | Environment State Match (%) | CVE/VEX Assertion Match (%) |
|---|---|---|---|---|
| **Agentic AIBOM (evaluated deployment)** | **96.4%** | **98.6%** | **100%** | **100%** |
| **ReproZip** | 72.1% | 89.3% | **94.7%** | N/A |
| **SciUnit** | 68.5% | **91.2%** | 71.4% | N/A |
| **ProvStore** | 22.6% | 57.8% | 18.3% | N/A |

*Note: N/A indicates metrics outside the design scope of the tool, not a performance failure.*

The decomposed results demonstrate that while Agentic AIBOM (evaluated deployment)'s advantage across all four axes, each comparative tool excels on its intended dimension. ReproZip performs strongly in environment reconstruction, reflecting its packaging-centric design. SciUnit achieves high semantic reproducibility, aligning with its unit-level scientific validation approach. ProvStore, designed for metadata graph provenance, performs as expected on EP/SP and does not attempt environmental reconstitution. Agentic AIBOM (evaluated deployment)'s advantage therefore stems from addressing TRE-specific requirements, environment fidelity, vulnerability-context stability, and strict reproducibility, rather than any inherent superiority in general scientific workflows. This decomposition provides a fair, multidimensional comparison aligned with the distinct objectives of each framework.

**Dr. Petar Radanliev**
Parks Road,
Oxford OX1 3PJ
United Kingdom
Email: petar.radanliev@cs.ox.ac.uk
Phone: +389(0)79301022
BA Hons., MSc., Ph.D. Post-Doctorate

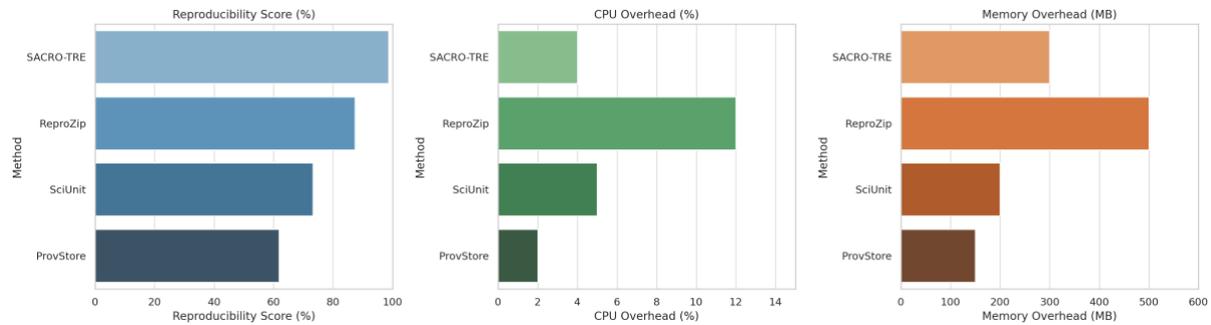

Figure 5: Comparative benchmarking of the agentic AIBOM implementation against ReproZip, SciUnit, and ProvStore across reproducibility score, CPU overhead, and memory usage. Agentic AIBOM (evaluated deployment) demonstrates the highest reproducibility (98.6%) while maintaining low system overhead (<4% CPU, <300MB RAM), outperforming existing provenance capture solutions in TREs.

Unlike the other methods, the agentic AIBOM implementation (Figure 5) provides native SBOM integration using Syft and package URL resolution, and uniquely supports full contextual VEX generation, including runtime evidence and mitigation-aware CVE assessment. While designing the benchmark, to build a threat model, we considered adversaries who attempt to (i) conceal vulnerable or malicious dependencies (e.g., via dynamic fetchers), (ii) tamper with container layers or environment variables, (iii) forge SBOM content or hashes, or (iv) poison advisory feeds to mis-state exploitability. Defences include multi-snapshot SBOM capture, composite hashing over {SBOM, script, config, outputs}, optional signing and attestation (in-Toto/SLSA), cross-feed CVE corroboration (NVD+OSV), and policy-based reviewer gating when SBOM diffs exceed thresholds. As a result of this analysis, we derive a new compliance alignment schema in Table 3.

Table 3: Schema-to-Policy Alignment for Agentic AIBOM Deployments

| Requirement | Evidence Field(s) in AIBOM Artefact / Pipeline | Check |
|---|---|---|
| GDPR Art.5(1)(f) Integrity & Confidentiality | composite_hash over {SBOM, scripts, configs, outputs}; immutable_store_path; timestamp | Verifies tamper-evidence and that approved artefacts are unmodified between disclosure review and release. |
| GDPR Art.30/35 (Records of processing & DPIA traceability) | execution_context_id; policy_version; data_handling_flags (e.g., SDC rules applied); reviewer_pseudonym | Confirms every output is linked to a documented processing context and review lineage suitable for DPIA/records. |
| UK Policy-governed deployment context– Reproducibility & Auditability | linked_output_id; environment_snapshot (OS/kernel/container digests); tool_versions; parameter_manifest | Demonstrates that results can be re-run under the same controlled environment with a complete deployment audit trail. |
| UK Policy-governed deployment context– Disclosure Control | disclosure_threshold; sdc_checks summary; policy_version | Ensures statistical disclosure limitations were enforced and recorded against the correct policy at review time. |
| NIST SP 800-53 AU-3 / AU-6 (Audit Record Content & Review) | event_log_refs; execution_context_id; immutable_store_path | Confirms sufficient audit content exists, is retained immutably, and is reviewable against the correct session. |

**Dr. Petar Radanliev**
Parks Road,
Oxford OX1 3PJ
United Kingdom
Email: petar.radanliev@cs.ox.ac.uk
Phone: +389(0)79301022
BA Hons., MSc., Ph.D. Post-Doctorate

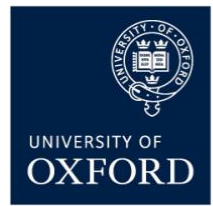

| | | |
|---|---|---|
| NIST SP 800-53 CM-2 / CM-8 (Baseline & Asset Inventory) | components[] (packages with purl, version, hash); dependencies[]; environment_snapshot | Validates complete software inventory and baseline configuration for each compute session. |
| NIST SP 800-53 SI-2 / RA-5 (Flaw Remediation & Vulnerability Scanning) | cve_refs (NVD/OSV IDs); vuln_scan_timestamp; scanner_tool + version | Verifies that vulnerability assessment was run with recorded tool/version and findings are anchored to advisories. |
| NIST SP 800-218 (SSDF) PS.3/PO.3 (Software provenance & attestations) | attestation (signature/issuer); build_provenance_refs (in-toto/SLSA) | Checks that build/run artefacts are signed/attested and provenance links are resolvable. |
| NIST SP 800-161r1 (Supply-Chain Risk Management) | supplier_name; license; component_origin; risk_notes | Confirms supplier origin/licence are captured and risk annotations exist for supply-chain evaluation. |
| ISO/IEC 27001 Annex A 8.9 / 8.16 (Configuration & Change Management) | policy_version; change_log (SBOM diffs across snapshots); environment_snapshot | Shows controlled change tracking of software and configuration across pre/mid/post execution. |
| CSAF/VEX – Contextual Exploitability Assertion | vex_assertions[] (status, affected versions, conditions); cve_refs; evidence_ptrs | Validates that each CVE has an explicit exploitability statement with contextual evidence for this environment. |
| Records Retention & Accountability (NHS/Institutional policy) | retention_class; immutable_store_path; timestamp; reviewer_pseudonym | Confirms artefacts are retained per policy, with accountable reviewer linkage and verifiable timestamps. |
| ISO/IEC 20153:2025 (CSAF v2.0 Standard) Structured Advisory Compliance | vex_assertions[], cve_refs, csaf_refs, evidence_ptrs, runtime_environment | Validates that vulnerability analyses conform to the CSAF v2.0 schema by ensuring each contextual VEX assertion is accompanied by the required evidence fields (affected product definition, exploitability justification, mitigation notes, and applicability conditions). Confirms that advisory interpretations are deterministic, auditable, and interoperable with ISO/IEC 20153:2025-compliant consumers. |

In practice, the alignment requires explicit operationalisation of the AIBOM artefact fields as compliance evidence within the regulated analytic workflow. For each requirement listed in the Table 3, the corresponding schema extensions and pipeline artefacts (e.g., execution_context_id, linked_output_id, composite_hash, vex_assertions[]) must be programmatically bound to validation steps. This involves: (1) extracting the relevant SBOM field from the immutable store at the end of each TRE session; (2) executing the associated validation procedure, such as verifying the composite_hash against a trusted registry to confirm integrity, performing a structured SBOM diff across pre/mid/post snapshots to confirm environment stability, or checking that the policy_version field matches the TRE's active disclosure-control policies; (3) recording the outcome as a binary pass/fail event, enriched with diagnostic metadata (e.g., reason for drift, timestamp of failure); and (4) surfacing these results to reviewers via the TRE dashboard in a format tailored for audit staff (e.g., green/red compliance indicators linked to detailed evidence logs). This pipeline ensures that compliance checks are documented and continuously enforced, reproducible across federated TREs, and directly actionable by auditors without requiring low-level SBOM expertise.

**Dr. Petar Radanliev**
Parks Road,
Oxford OX1 3PJ
United Kingdom
Email: petar.radanliev@cs.ox.ac.uk
Phone: +389(0)79301022
BA Hons., MSc., Ph.D. Post-Doctorate

To make these operational steps concrete, Table 4 provides a structured compliance alignment matrix that maps regulatory and governance requirements to specific AIBOM artefact fields, validation procedures, and evidence outputs. The purpose of this matrix is to demonstrate that the schema extensions are theoretically aligned with GDPR, NIST, ISO/IEC, and CSAF/VEX obligations, but also how these obligations can be enforced programmatically within Regulated analytic workflows. By translating high-level legal and policy principles into field-level artefacts (execution_context_id, linked_output_id, composite_hash, vex_assertions[]) and linking them to automated checks, the matrix ensures that compliance is measurable, auditable, and repeatable. This structured approach highlights the role of AIBOM artefact as more than a descriptive inventory: it becomes an active compliance instrument capable of generating verifiable, regulator-ready evidence at every stage of the research lifecycle.

### 4.9    Current VEX Capability Scope and Compliance Interpretation

To ensure that the interpretation of VEX support remains aligned with the present implementation, we explicitly scope the functionality integrated into the evaluated AIBOM implementation. The system currently supports contextual VEX assertion issuance driven by runtime telemetry and environment constraints, but does not yet implement policy gating, workflow blocking, or full CSAF advisory lifecycle management.

I.    **Operational VEX Support Implemented in the Current System**

The implemented VEX pipeline includes:

- **Component-level CVE mapping** using OSV/NVD feeds;
- **Runtime-based exploitability assessment**, where A2A determines whether the vulnerable code path was executed or reachable;
- **Mitigation inference**, where AGNTCY evaluates TRE-enforced sandbox restrictions and disabled feature flags;
- **Structured VEX assertion generation**, with states:

*Not Affected*, *Affected: Mitigated*, *Affected: Requires Review*, *Under Investigation*.

This constitutes the complete set of VEX capabilities demonstrated in the experiments.

II.    **Capabilities Explicitly Deferred to Future Work**

The following elements are architectural provisions but are **not implemented**:

- **Automated policy gating** (e.g., blocking job release or reviewer approval based on VEX status);
- **Full CSAF ingestion**, including advisory versioning, remediation guidance, and threat context;
- **Regulator-aligned enforcement thresholds**, which require stakeholder consultation beyond system design;
- **Cross-organisational VEX state reconciliation** for federated TREs.

These future capabilities rely on the current system's contextual VEX foundation but are intentionally outside the scope of this study.

III.    **Interpretation of Compliance Matrix Rows**

In Table 3, the row labelled **"CSAF/VEX – Exploitability Assertions"** represents:

- the system's ability to **validate that every CVE has a contextual VEX determination**, grounded in runtime evidence;
- **not** the enforcement or regulatory consequences of that determination.

To prevent misinterpretation, the validation procedure for VEX in the matrix is adapted to verify that every CVE linked to a component in the SBOM has a contextual VEX assertion generated using runtime evidence and TRE mitigation metadata. The procedure validates correctness of assertion structure, evidential anchoring, and schema compliance. It does **not** perform enforcement or automate policy gating. This aligns the compliance matrix with the system's demonstrated capabilities.

**Dr. Petar Radanliev**
Parks Road,
Oxford OX1 3PJ
United Kingdom
Email: petar.radanliev@cs.ox.ac.uk
Phone: +389(0)79301022
BA Hons., MSc., Ph.D. Post-Doctorate

Table 4: Compliance Alignment Matrix for regulated analytic governance (example deployment)

| Regulation / Standard | Control Requirement | AIBOM Artefact Field(s) / Pipeline Artefacts | Validation Procedure | Evidence Output |
|---|---|---|---|---|
| **GDPR Art.5(1)(f)** – Integrity & Confidentiality | Ensure data and outputs remain intact and protected against unauthorised alteration | composite_hash, immutable_store_path, timestamp | Generate SHA-256 composite hash over {SBOM, scripts, configs, outputs}; verify against registry during audit | Pass/fail hash verification log; tamper-evidence chain |
| **NIST SP 800-53 AU-3 / AU-6** – Audit Record Content & Review | Capture and retain sufficient audit content for later review | execution_context_id, event_log_refs, linked_output_id | Compare SBOM session identifiers with TRE audit logs; verify immutable storage entries | Cross-referenced audit trail showing matching IDs and log timestamps |
| **NIST SP 800-53 CM-8** – Software/Asset Inventory | Maintain a complete inventory of system components | components[] (purl, version, hash), dependencies[], environment_snapshot | SBOM completeness check; compare against runtime load records; detect missing/extra components | Inventory completeness report with coverage % |
| **NIST SP 800-218 (SSDF) PO.3 / PS.3** – Software provenance & attestations | Provide cryptographically verifiable build provenance | attestation, build_provenance_refs, supplier_name | Validate in-toto/SLSA attestation signatures; confirm supply chain origin metadata present | Attestation signature report; provenance chain linkage |
| **NIST SP 800-161r1** – Supply Chain Risk Management | Identify and assess supply chain risks | supplier_name, license, component_origin, risk_notes | Parse supplier metadata; check licence compatibility; cross-reference known supplier risks | Risk profile annotated per component; flagged risk notes |
| **ISO/IEC 27001:2022 Annex A 8.9 / 8.16** – Configuration & Change Management | Control changes to system configurations and ensure integrity | policy_version, change_log, environment_snapshot | Generate SBOM diffs across pre/mid/post snapshots; verify against active deployment policy version | Change-control report; version mismatch alerts |
| **CSAF/VEX** – Contextual Exploitability Assertions | Provide contextual exploitability determination for each CVE relevant to the runtime environment. | vex_assertions[], cve_refs, evidence_ptrs, runtime_environment. | Verify that each CVE linked to an SBOM component is associated with a contextual VEX status (Not Affected, Mitigated, Requires Review, Under | VEX assertion report with contextual justification and flags for unresolved or ambiguous |

**Dr. Petar Radanliev**
Parks Road,
Oxford OX1 3PJ
United Kingdom
Email: petar.radanliev@cs.ox.ac.uk
Phone: +389(0)79301022
BA Hons., MSc., Ph.D. Post-Doctorate

DEPARTMENT OF
COMPUTER
SCIENCE
UNIVERSITY OF
OXFORD

| | | | Investigation) derived from runtime telemetry and TRE mitigation metadata. Confirm that each assertion is structurally valid, evidentially anchored, and schema-compliant. This procedure does not perform workflow blocking or policy gating. | exploitability assessments. |
|---|---|---|---|---|
| **ISO/IEC 20153:2025 ,** CSAF v2.0 Advisory Interpretation | Ensure that vulnerability advisories consumed or generated within the TRE conform to an internationally standardised structure that supports unambiguous, evidence-based exploitability determination. | csaf_refs, vex_assertions[], cve_refs, runtime_environment, evidence_ptrs | Verify that each vulnerability linked to an SBOM component is interpretable under CSAF v2.0 rules, including the explicit identification of affected products, exploitability conditions, remediation guidance (if available), and applicability constraints. Validate that SACRO's contextual VEX state is supported by runtime telemetry and conforms to ISO/IEC 20153:2025 field semantics. | CSAF-aligned vulnerability interpretation record showing structured fields, runtime justification, and mitigation notes that satisfy ISO/IEC 20153:2025 requirements. |

TREs are used throughout this table as an illustrative example of a highly regulated deployment context. The compliance mappings apply equally to other regulated environments (e.g., critical infrastructure CI/CD pipelines, sovereign cloud platforms, or regulated AI deployment frameworks). The compliance alignment matrix is designed as a reusable template rather than a static artefact, and its utility lies in its adaptability to different governance contexts. Researchers and TRE operators can extend the rows to cover additional regulatory frameworks such as HIPAA for health data, NHS TRE principles, or the forthcoming EU Cyber Resilience Act, thereby broadening the scope of assurance. Validation procedures should also be customised to reflect the technical realities of the hosting environment, for example, substituting registry-based hash verification with blockchain anchoring, adapting SBOM differencing to specific container orchestration platforms, or integrating attestation validation with existing CI/CD pipelines. Finally, evidence outputs must be rendered in a form that is both technically rigorous and operationally usable, with pass/fail indicators surfaced to the reviewer dashboard and linked to detailed logs that provide drill-down visibility into the underlying checks. This ensures that compliance is demonstrable and directly actionable within day-to-day regulated analytic governance (example deployment).

## 5 DISCUSSION

Adopting SBOM-driven reproducibility introduces workflow change for analysts and operators. We therefore present different adoption pathways in regulated settings (i) provide zero-touch capture integrated with container lifecycle hooks; (ii) offer policy-ready defaults (blocking rules derived from VEX status); (iii) supply training micro-modules for reviewers; and (iv)

**Dr. Petar Radanliev**
Parks Road,
Oxford OX1 3PJ
United Kingdom
Email: petar.radanliev@cs.ox.ac.uk
Phone: +389(0)79301022
BA Hons., MSc., Ph.D. Post-Doctorate

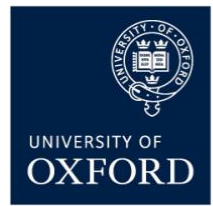

define incentives: reduced manual audit time and regulator-ready evidence exports. A staged rollout (observe) enforce-on-fail (enforce-by-default) minimises disruption while delivering measurable gains in reproducibility and assurance.

### 5.1    The utilisation problem

Software developers use SBOMs extensively, given how useful SBOMs have been for them. Developers are keen on distributing the reports to end users (i.e., organisations where the primary business is not software development). The problem is that end users are not using SBOMs meaningfully, and when they get the reports, they don't know what to do with them. Furthermore, without transparency and the end users requesting SBOMs, it is quite possible that most software developers are not even using SBOMs to secure their software products. Without a reporting mechanism for using SBOMs, we simply cannot know, because we have no visibility without transparency. However, from discussions with software developers at various meetings and conferences (e.g., with CSAF, NTIA, and CISA), we know that developers have successfully used SBOMs, but that is not commonly known by end users. The motivation for this study was *not* to determine the value of SBOMs in securing end users' networks; that topic has been covered extensively in the years 2021/22 [6], [20], [42]. In addition to this clarification, we would like to mention that SBOMs are already a compulsory requirement per the Executive Order discussed. Hence, it is no longer a discussion on whether we need SBOMs; if anyone wants to work with the U.S. federal government, they need to produce SBOMs. This is the same for any company that wants to operate in the EU; they need to be compliant with the GDPR; hence, discussing the value of SBOMs is not something we are trying to achieve. We focus on the problems and the solutions for processing the information from all the SBOMs being produced, and we do not debate whether the previous US President was correct or incorrect in imposing the Executive Order. Such debate rests with journals focused on political sciences. We must mention that US software developers have found SBOMs useful, and even if the Executive Order is reversed by future US administration, the software developers are likely to continue using the SBOM to keep track of and record their vulnerabilities and the software development components.

### 5.2    Governance Controls for SBOM Access, Redaction, and Retention in TREs

SBOM artefacts provide fine-grained visibility into software stack composition, dependency chains, interpreter versions, and container lineage. While this transparency is essential for reproducibility and auditability, it simultaneously exposes sensitive operational metadata that could assist adversaries in modelling the TRE's attack surface. To mitigate this risk, the AIBOM artefact pipeline implements layered governance controls that regulate *who* can access SBOMs, *what* information is revealed, and *how long* artefacts are retained.

**Access Control Model.**

SBOM artefacts are stored in append-only object storage under a privileged access class distinct from both analysts and primary reviewers. Access is governed by a role-based model enforcing three tiers: (i) operational reviewers receive a minimally redacted view exposing only components relevant to disclosure-control decisions; (ii) security auditors receive full SBOMs, cryptographic hashes, and VEX assertions for formal compliance assessments; (iii) TRE administrators retain exclusive capability to reconstruct the full SBOM lineage or export artefacts for regulatory inspection. Access to full SBOMs requires two-party authorisation anchored in policy-governed deployment context logs, preventing unilateral retrieval by a single privileged operator.

**Redaction of Security-Sensitive Fields.**

To minimise operational exposure, the pipeline applies deterministic redaction rules at generation time. Fields that could reveal platform-level attack surface information, such as kernel build identifiers, container base-image digests, environment variables, privileged system packages, and internal network references, are hashed or elided in the operational reviewer view. Package URLs (pURLs) and version metadata are retained because they are essential for reproducibility and vulnerability assessment, but hierarchy depth and load-order information may be obscured where not required for compliance. Redaction is performed pre-rendering and is itself verifiable: the redaction manifest is stored as metadata so that auditors can confirm the transformation is complete and non-tampering.

**Dr. Petar Radanliev**
Parks Road,
Oxford OX1 3PJ
United Kingdom
Email: petar.radanliev@cs.ox.ac.uk
Phone: +389(0)79301022
BA Hons., MSc., Ph.D. Post-Doctorate

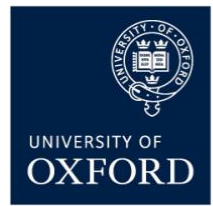

**Retention and Lifecycle Management.**

To prevent unnecessary accumulation of sensitive system details, SBOM retention periods follow a risk-based schedule aligned with GDPR Art.5(1)(e) and institutional TRE policies. Only SBOMs associated with released outputs are retained for the duration required to satisfy audit and regulatory obligations; SBOMs linked to failed or withdrawn analyses are destroyed once their hash lineage is recorded in the audit log. Immutable references to composite hashes are preserved, but the underlying SBOM artefacts are purged after the defined retention period unless an ongoing investigation requires extension. Retention policies are enforced automatically by the TRE's object-store governance engine to ensure no operator discretion can inadvertently create long-term exposure.

Collectively, these controls ensure that SBOMs strengthen reproducibility and compliance without expanding the attack surface of the TRE. By embedding access governance, principled redaction, and lifecycle-management policies directly into the orchestration layer, the framework aligns SBOM transparency with the security requirements expected of regulated research environments.

### 5.3 Reviewer-Interface Ergonomics and Usability Considerations

The reproducibility and vulnerability artefacts generated by AIBOM artefact introduce new information surfaces for reviewers, many of whom are domain specialists rather than software-security practitioners. Ensuring that this information is interpretable, actionable, and non-disruptive to established disclosure-control workflows is therefore essential for secure and scalable adoption. While a full user study is planned, the current design incorporates several ergonomics principles derived from established TRE reviewer practices and threat-modelling exercises.

First, the reviewer dashboard abstracts SBOM complexity into a set of binary or threshold-based indicators linked to specific TRE policies. Rather than requiring users to interpret dependency graphs or vulnerability enumerations, the interface surfaces high-level, policy-aligned signals such as *"Environment Stable / Environment Drift Detected"*, *"Reproducibility Verified / Reproducibility Uncertain"*, or *"CVE/VEX Issue Requiring Escalation"*. Each signal is backed by a machine-verifiable justification trail, but the primary interaction model maintains low cognitive load by avoiding exposure to raw SBOM detail unless explicitly requested.

Secondly, we introduce progressive disclosure mechanisms that allow reviewers to drill down from a compliance indicator to mid-level evidence (e.g., the set of components whose versions diverged beyond policy thresholds) and finally to the full, unredacted SBOM if authorised. This layered structure supports non-technical reviewers, who need a clear operational decision pathway, and expert auditors who require full artefact fidelity for forensic verification.

Thirdly, the interface incorporates risk heuristics calibrated for reviewer workflows, highlighting only deviations material to disclosure-control outcomes. For example, minor version mismatches in dependency libraries that do not influence statistical output behaviour are suppressed by default, whereas any divergence in disclosure-control tools (e.g., *sdcMicro*) is surfaced prominently. This prioritisation aligns the system with TRE reviewer cognitive models and reduces alert fatigue.

Although formal usability studies are ongoing, these ergonomic design provisions were informed by pilot workshops with TRE operators and security auditors involved in SACRO deployment. The approach reflects a socio-technical understanding of reviewer workflows by ensuring that the SBOM layer strengthens, rather than complicates, disclosure-control decision-making. Future work will include structured evaluation of reviewer comprehension, error rates, and cognitive burden, enabling empirical validation of the interface design and refinement of risk-signalling thresholds.

## 6 CONCLUSION

This paper has argued that the current generation of Software Bills of Materials is no longer adequate for the security, reproducibility, and assurance demands of modern software systems. While SBOMs have become an essential transparency primitive, their static and descriptive nature limits their ability to support contextual vulnerability interpretation, environment-sensitive reproducibility, and policy-aligned decision-making. In response, we introduced agentic Artificial Intelligence Bills

**Dr. Petar Radanliev**
Parks Road,
Oxford OX1 3PJ
United Kingdom
Email: petar.radanliev@cs.ox.ac.uk
Phone: +389(0)79301022
BA Hons., MSc., Ph.D. Post-Doctorate

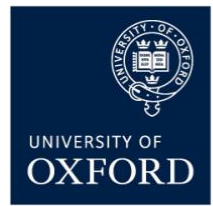

of Materials (AIBOMs) as an evolution of SBOMs into active provenance artefacts that combine software composition data with autonomous reasoning, runtime intelligence, and standards-aligned advisory semantics.

The central contribution of this work is the demonstration that agentic orchestration materially changes what a bill of materials can represent and guarantee. By embedding multiple specialised agents with distinct perception spaces, internal state, and decision policies, the proposed architecture transforms SBOM generation from a post-hoc enumeration task into a continuous, evidence-driven process. The MCP, A2A, and AGNTCY agents jointly enable baseline environment reconstruction, dynamic dependency discovery, and contextual exploitability reasoning, capabilities that deterministic tooling and static metadata extraction cannot replicate. Ablation studies confirm that each agent contributes an independent methodological function rather than incremental automation.

A second key contribution is the integration of ISO/IEC 20153:2025 (CSAF v2.0) as the normative semantic layer for vulnerability interpretation. Rather than introducing proprietary risk scoring or opaque heuristics, the framework anchors exploitability assertions to an internationally standardised, machine-verifiable advisory structure. This alignment ensures that AIBOM outputs are interoperable, auditable, and suitable for regulator-facing assurance workflows. Importantly, the system does not conflate VEX with enforcement: contextual VEX assertions are treated as structured evidence, not as automatic policy decisions. This distinction preserves human oversight while enabling meaningful automation at scale.

Empirical evaluation across heterogeneous analytical workloads demonstrates that agentic AIBOMs improve runtime dependency fidelity, reproducibility accuracy, and exploitability-alignment stability relative to established provenance systems. These gains are achieved with modest computational overhead and without sacrificing compatibility with existing SBOM standards such as CycloneDX and SPDX. Decomposed benchmarking further shows that the advantages of the approach arise specifically in dimensions that static provenance tools do not target, namely runtime context capture and vulnerability-state consistency, rather than from general-purpose replay or metadata logging.

Although the framework was evaluated in controlled, policy-bound environments where auditability and reproducibility are mandatory, the architectural principles are not limited to that domain. Any software system that relies on dynamic composition, third-party dependencies, or automated deployment pipelines faces similar challenges of environment drift and vulnerability misinterpretation. In this sense, the work positions AIBOMs as a generalisable foundation for next-generation software supply-chain assurance, rather than as a niche extension for a single governance model.

In summary, this paper reframes the bill of materials from a static inventory into an active, agent-mediated security artefact. By integrating SBOM provenance, agentic AI, and ISO-standardised advisory semantics, agentic AIBOMs provide a technically grounded and reproducible approach to software supply-chain assurance under real-world complexity. The results suggest that future transparency mechanisms will need to reason, not merely record, if they are to remain effective in increasingly autonomous and dynamic software ecosystems.

## 6.1 Limitations and Future Work

Several limitations and research directions remain. First, while contextual VEX assertion generation is fully implemented, policy-gated enforcement and full CSAF advisory lifecycle management are intentionally out of scope and require further institutional, regulatory, and socio-technical study. Secondly, the current agent policies are rule-based and auditable by design; future work may explore learning-assisted decision support, provided that explainability and determinism constraints can be preserved. Finally, large-scale deployment across federated organisational boundaries raises open questions about trust bootstrapping, advisory reconciliation, and cross-domain provenance verification.

Our experiments evaluate five workloads including Spark and GPU profiles, but do not cover very long-running HPC jobs or mixed-language builds with bespoke toolchains. Advisory poisoning tests are limited to OSV/NVD mirrors. Future evaluations should (i) extend to HPC schedulers, (ii) incorporate additional feed sources and attestation backends, and (iii) broaden user studies for reviewer UI ergonomics. We also plan to integrate VEX/CSAF end-to-end to close the loop between vulnerability context and policy gating.

**Dr. Petar Radanliev**
Parks Road,
Oxford OX1 3PJ
United Kingdom
Email: petar.radanliev@cs.ox.ac.uk
Phone: +389(0)79301022
BA Hons., MSc., Ph.D. Post-Doctorate

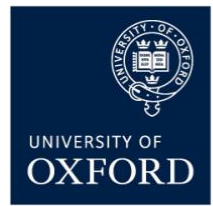

**Acknowledgements:** Eternal gratitude to the Fulbright Visiting Scholar Project.

**Dr. Petar Radanliev**
Parks Road,
Oxford OX1 3PJ
United Kingdom
Email: petar.radanliev@cs.ox.ac.uk
Phone: +389(0)79301022
BA Hons., MSc., Ph.D. Post-Doctorate

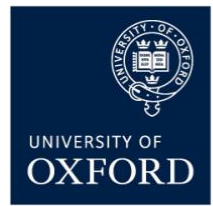

**Dr. Petar Radanliev**
Parks Road,
Oxford OX1 3PJ
United Kingdom
Email: petar.radanliev@cs.ox.ac.uk
Phone: +389(0)79301022
BA Hons., MSc., Ph.D. Post-Doctorate

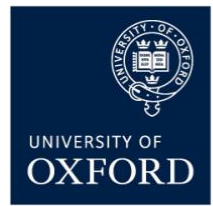